\documentclass[a4paper,fleqn,usenatbib]{mnras}

\usepackage{newtxtext,newtxmath} 
\usepackage[T1]{fontenc} 
\usepackage{ae,aecompl}

\usepackage{graphicx}	% Including figure files 
\usepackage{amsmath}	% Advanced maths commands 
\usepackage{amssymb}	% Extra maths symbols

\newcommand{\elecd}{$n_{\rm e}$} 
\newcommand{\elect}{$T_{\rm e}$} 
 
\newcommand{\hb}{H$\beta$} 
\newcommand{\ha}{H$\alpha$} 
 
\newcommand{\foi}{[O\thinspace{\sc i}]} 
\newcommand{\foii}{[O\thinspace{\sc ii}]} 
\newcommand{\foiii}{[O\thinspace{\sc iii}]} 
\newcommand{\fsii}{[S\thinspace{\sc ii}]} 
\newcommand{\fsiii}{[S\thinspace{\sc iii}]} 
\newcommand{\fni}{[N\thinspace{\sc i}]} 
\newcommand{\fnii}{[N\thinspace{\sc ii}]} 
\newcommand{\fniii}{[N\thinspace{\sc iii}]} 
 
\newcommand{\fcliii}{[Cl\thinspace{\sc iii}]}

\newcommand{\fneiii}{[Ne\thinspace{\sc iii}]} 
 
\newcommand{\ffeii}{[Fe\thinspace{\sc ii}]} 
\newcommand{\ffeiii}{[Fe\thinspace{\sc iii}]}

\newcommand{\oi}{O\thinspace{\sc i}} 
\newcommand{\oii}{O\thinspace{\sc ii}} 
 
\newcommand{\cii}{C\thinspace{\sc ii}}

\newcommand{\fariii}{[Ar\thinspace{\sc iii}]}

\newcommand{\hi}{H\,{\sc i}} 
\newcommand{\hii}{H\thinspace{\sc ii}} 
 
\newcommand{\hei}{He\thinspace{\sc i}} 
\newcommand{\heii}{He\thinspace{\sc ii}}

\newcommand{\ts}{\emph{$t^2$}}

\newcommand\ionic[2]{${\rm #1^{#2}}$}           % ej. \ionic{O}{+} --> O^+

\newcommand{\cmc}{{\rm cm$^{-3}$}}

\title[Revisiting the Galactic N and O abundance gradients]{Revisiting the radial abundance gradients of nitrogen and oxygen of the Milky Way}

\author[C. Esteban and J. Garc{\'{\i}}a-Rojas]{ C. Esteban,$^{1, 2}$\thanks{E-mail: cel@iac.es (CE)} and J. Garc{\'{\i}}a-Rojas$^{1, 2}$\thanks{E-mail: jogarcia@iac.e (JGR)} 
\\ 
$^{1}$Instituto de Astrof\'isica de Canarias, E-38200 La Laguna, Tenerife, Spain\\ 
$^{2}$Departamento de Astrof\'isica, Universidad de La Laguna, E-38206, La Laguna, Tenerife, Spain\\ }

\date{Accepted XXX. Received YYY; in original form ZZZ}

\pubyear{2018}

\begin{document} 
\label{firstpage} 
\pagerange{\pageref{firstpage}--\pageref{lastpage}} 
\maketitle


\begin{abstract} We present spectra obtained with the 10.4 m {\it Gran Telescopio Canarias} telescope of 13 Galactic {\hii} regions, most of them of very low ionisation degree. The objects are located along the Galactic disc, with $R_{\rm G}$ from 5.7 to 16.1 kpc. We determine \elect({\fnii}) for all of them. We obtain -- for the first time -- a radial abundance gradient of N that is independent on the ionisation correction factor. The radial distribution of the N/O ratio is almost flat, indicating that the bulk of N is not formed by standard secondary processes. We have made a reassessment of the radial O abundance gradient combining our results with previous similar ones by \citet{estebanetal17}; producing a homogeneous dataset of 35 {\hii} regions with direct determinations of the electron temperature. We report the possible presence of a flattening or drop of the O abundance in the inner part of the Galactic disc. This result confirms previous findings from metallicity distributions based on Cepheids and red giants. Finally, we find that the scatter of the N and O abundances of {\hii} regions with respect to the gradient fittings is not substantially larger than the observational uncertainties, indicating that both chemical elements seem to be well mixed in the interstellar gas at a given distance along the Galactic disc 
\end{abstract}



%%%%%%%%%%%%%%%%%%%%%%%%%%%%%%%%%%%%%%%%%%%%%%%%%%

%%%%%%%%%%%%%%%%% BODY OF PAPER %%%%%%%%%%%%%%%%%%

%%%%%%%%%%% 
\section{Introduction} 
\label{sec:intro} 
%%%%%%%%%%%

The determination of the radial distributions of chemical abundances along galactic discs -- radial gradients -- are essential observational constraints for chemical evolution models. Radial gradients reflect the star formation history and the effects of gas flows and other processes over the chemical composition of the galaxies. {\hii} regions trace the present-day composition of the gas phase of the interstellar medium. They have been largely used to determine the radial abundance gradients of several elements, especially of O, the proxy of metallicity in the analysis of ionised nebula. In these objects, the O abundance can be derived simply adding the ionic abundances of O$^+$ and O$^{2+}$, that can be obtained from the intensity of bright optical collisionally excited lines (hereafter CELs). However, N does not show observable CELs of N$^{2+}$ in the optical spectral range, and we have to assume an ionisation correction factor (ICF) to estimate the total N abundance. There are several schemes of ICF(N$^+$) available in the literature based on the similarity of the ionisation potential of O and N or on photoionisation models. The dependence on the assumption of an ICF(N$^+$) scheme is an important drawback to derive the N abundance because N$^{2+}$ is usually the dominant ionisation stage of this element in {\hii} regions.

The evolution of N in galaxies is complex. It can be produced in both massive and intermediate-mass stars and can have a primary or secondary character depending on the origin of the C and O atoms from which the bulk of N is originated. A precise determination of the radial gradient of N across the disc of the Milky Way is essential for better understanding the origin of this element. The radial distribution of the N/O ratio gives information about the primary or secondary character of N. Determinations of the radial gradient of N/O based on optical or far infrared (FIR) spectra of {\hii} regions give different results. Based on optical spectra, \citet{shaveretal83} obtain an almost flat radial gradient of N/O while \citet{rudolphetal06}, using FIR observations, find an apparently negative one. Therefore, N seems to be primary or secondary depending on the kinds of observations we are using, which is clearly an inconsistent result. The N abundance determinations based on FIR spectra have also to rely on the assumption of an ICF but in this case for the contribution of N$^+$ because only CELs of N$^{2+}$ are observable in the FIR.

The nucleosynthesis of O is simpler than that of N. It is produced mostly by massive, short-lived stars and the N enrichment of the ISM should be delayed with respect to the O enrichment. There are 
many determinations of the radial O abundance gradient of the Galactic disc based on {\hii} regions observations \citep[e.g.][among others]{shaveretal83, deharvengetal00, rudolphetal06, balseretal11, estebanetal17}. The most recent determinations indicate that the gradient slope is between $-$0.040 and $-$0.060 dex kpc$^{-1}$. Despite some authors claimed that the radial O 
abundance gradient may flatten out at the outer parts of the Milky Way \citep[e. g.][]{fichsilkey91, vilchezesteban96, macieletal06}, the recent work by \cite{estebanetal17} has confirmed the absence 
of such flattening, indicating that the slope of the gradient remains constant at least up to $R_{\rm G}$ $\sim$ 17 kpc. On the other hand, some works on metallicity gradients based on Cepheids 
and red giants observations have found indications of a flattening of the gradients in the inner Galactic disc, at $R_{\rm G}$ $<$ 5-6 kpc \citep{haydenetal14, martinetal15, andrievskyetal16}. An   
indication of the presence of such feature can be noted in the {\hii} region abundance gradient determined by \citet{estebanetal17}, although they did not report the fact because of the limited number of objects.

Until recently, the number of Galactic {\hii} regions with direct determination of {\elect} from optical spectra has been rather limited. \citet{peimbertetal78} determined the O abundance of a small number of objects and obtaining {\elect} from the same optical spectra. A somewhat larger number of {\hii} regions -- eight -- were used by \citet{estebanetal04}. One of the most cited papers on the Galactic abundance gradient, \citet{shaveretal83}, used optical spectra of 33 {\hii} regions, but they determined {\elect} from radio observations. Optical and radio observations were not cospatial and the aperture sizes of both kinds of data were very different. \citet{deharvengetal00} also combined non-cospatial optical spectra and radio observations to derive {\elect} for most of the 34 {\hii} regions of their sample. The studies of the radial abundance gradients based on FIR observations as that by \citet{rudolphetal06}, for example, combine measurements of CELs from FIR spectra and {\elect} determined from radio observations with also different apertures. In a recent paper, \citet{estebanetal17} calculated the Galactic radial O abundance gradient based on optical spectra of 21 {\hii} regions with direct determination of {\elect} from the same spectra. This could be achieved due to the use of very deep spectra obtained with 8-10 m telescopes that permit to measure the faint auroral CELs -- which intensities are strongly dependent on {\elect} -- in many very low surface brightness nebulae.

In this paper, we present very deep optical spectra of 13 {\hii} regions of low or very low ionisation degree with the aim of determining the N abundance eliminating as much as possible the effect of assuming an ICF and to increase the number of objects with O abundances calculated from cospatial determinations of {\elect}. The structure of this paper is as follows. In Section~\ref{sec:sample}, we describe the sample selection and in Section~\ref{sec:obs} the observations and data reduction procedure. In Section~\ref{sec:lines}, we describe the emission-line measurements and the reddening correction. In Section~\ref{sec:results} we present the physical conditions and ionic and total abundances determined for the sample objects. In Section~\ref{sec:Ngradient}, we present and discuss our determinations of the Galactic radial N abundance gradient. In Section~\ref{sec:icfN} we propose a new empirical ICF(N$^+$). In sections~\ref{sec:NOgradient} and \ref{sec:Ogradient} we present and discuss our determinations of the radial N/O and O gradients, respectively. Finally, in Section~\ref{sec:conclusions} we summarize our main conclusions.

%%%%%%%%%%%%%%% 
\begin{figure*} 
\centering \includegraphics[scale=0.15]{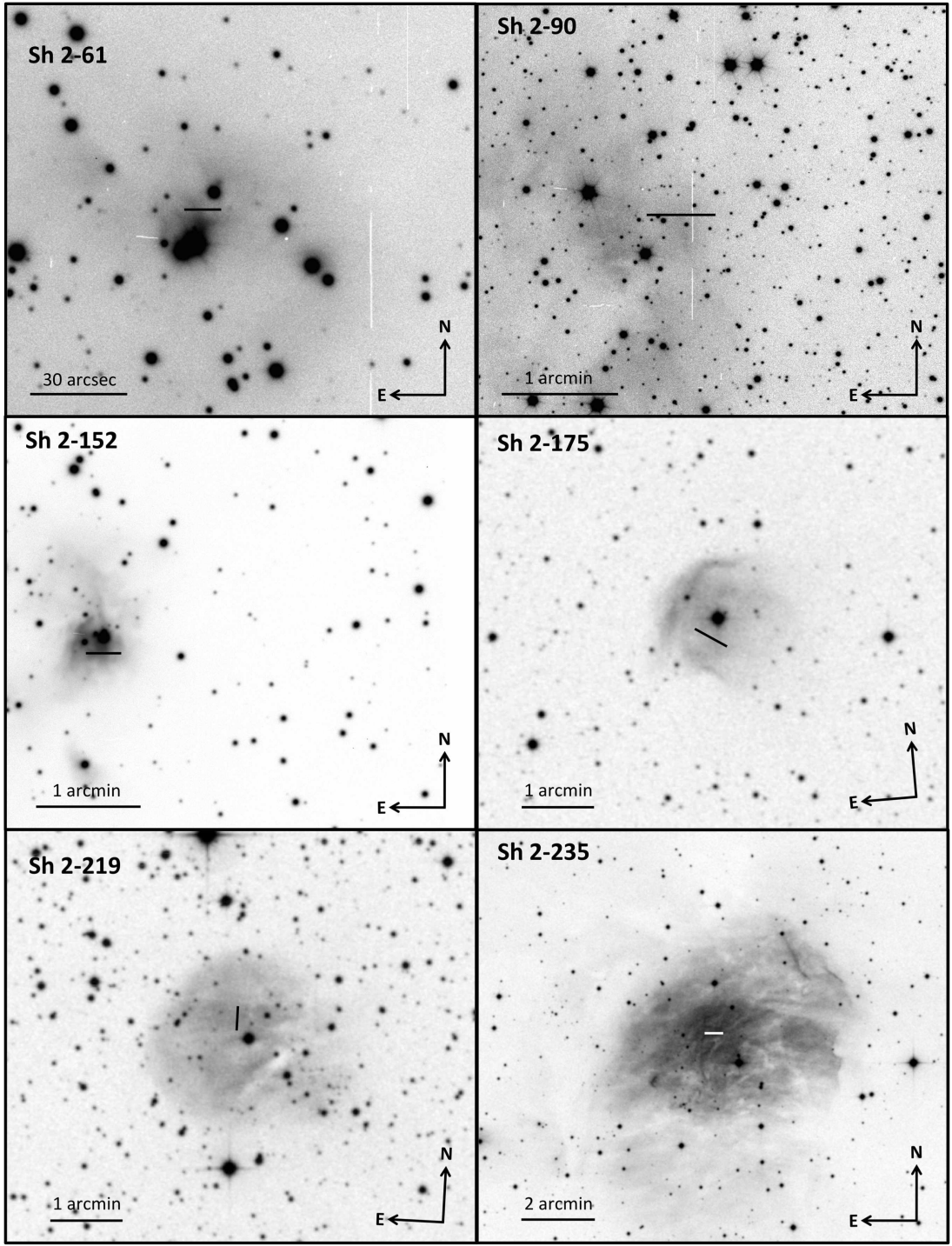} 
\caption{Finding charts of Sh~2-61, Sh~2-90, Sh~2-152, Sh~2-175, Sh~2-219 and Sh~2-235 indicating the position and length of the areas extracted for spectroscopical analysis (black/white lines). The charts of Sh~2-61, Sh~2-90 and Sh~2-152 are acquisition images taken with a {\it g} filter at the GTC. The charts of Sh~2-175, Sh~2-219 and Sh~2-235 correspond to red images of the Second Digitized Sky Survey (DSS-2) obtained from the Aladin interactive sky atlas \citep{bonnareletal00}.} \label{fig:slits_1} \end{figure*} 
%%%%%%%%%%%%%%%

%%%%%%%%%%%%% 
\begin{table*} 
\centering \caption{Data of the sample objects and their observations.} 
\label{tab:journal} 
\begin{tabular}{lccccccc} 
\hline 
& R.A.$^{\rm a}$ & Decl.$^{\rm a}$ & $R_\mathrm{G}^{\rm b}$ & PA & Extracted area & &  \\ 
{\hii} region & (J2000) & (J2000) &  (kpc) &  ($^\circ$) & ($\mathrm{arcsec^2}$) &Airmass & $S$(H$\alpha$)$^{\rm c}$\\ 
\hline 
Sh 2-61 & 18:33:21.04 & $-$04:57:55.7 & 5.7 $\pm$ 0.3 & 90 & 11.4 $\times$ 0.8 & 1.39 & 2.30 \\ 
Sh 2-90 & 19:49:09.33 & 26:51:41.4 & 7.4 $\pm$ 0.3 & 90 & 25.4 $\times$ 0.8 & 1.07 & 0.86 \\ 
Sh 2-152 & 22:58:40.80 & 58:46:53.8 & 10.3 $\pm$ 1.0 & 90 & 20.3 $\times$ 0.8 & 1.16 & 10.30 \\ 
Sh 2-175 & 00:27:47.93 & 64:42:00.3 & 10.2 $\pm$ 1.0 & 67 & 30.5 $\times$ 0.8 & 1.24 & 0.27 \\ 
Sh 2-219 & 04:56:11.69 & 47:23:49.4 & 12.3 $\pm$ 0.6 & 0 & 20.3 $\times$ 0.8 & 1.21 & 0.40 \\ 
Sh 2-235 & 05:41:03.12 & 35:51:33.1 & 9.6 $\pm$ 0.3 & 90 & 30.5 $\times$ 0.8 & 1.05 & 1.55 \\ 
Sh 2-237 & 05:31:26.10 & 34:15:06.2 & 10.2 $\pm$ 1.2 & 60 & 10.2 $\times$ 0.8 & 1.13 & 1.80 \\ 
Sh 2-257 & 06:12:43.99 & 17:58:52.4 & 10.1 $\pm$ 0.5 & 25 & 25.4 $\times$ 0.8 & 1.12 & 0.87 \\ 
Sh 2-266 & 06:18:46.14 & 15:17:27.6 & 14.5 $\pm$ 1.4 & 33 & 5.1 $\times$ 0.8 & 1.56 & 0.70\\ 
Sh 2-270 & 06:10:12.75 & 12:48:44.6 & 16.1 $\pm$ 1.4 & 66 & 5.1 $\times$ 0.8 & 1.07 & 0.66 \\ 
Sh 2-271 & 06:14:53.89 & 12:21:11.2 & 12.0 $\pm$ 0.7 & 43 & 33.0 $\times$ 0.8 & 1.06 & 0.85 \\ 
Sh 2-285 & 06:55:16.49 & $-$00:31:01.9 & 13.8 $\pm$ 0.2 & 90 & 21.6 $\times$ 0.8 & 1.28 & 0.55 \\ 
Sh 2-297 & 07:05:16.03 & $-$12:19:45.4 & 8.7 $\pm$ 0.1 & 104 & 41.9 $\times$ 0.8 & 1.37 & 2.06 \\ 
\hline 
\end{tabular} 
\begin{description} 
\item[$^{\rm a}$] Coordinates of the centre of the extracted aperture. 
\item[$^{\rm b}$] Galactocentric distances assuming the Sun at 8 kpc. 
\item[$^{\rm c}$] Measured mean H$\alpha$ surface brightness in the areas extracted for spectroscopical analysis in units of 10$^{-14}$ erg cm$^{-2}$ s$^{-1}$ arcsec$^{-2}$.
\end{description} 
\end{table*} 
%%%%%%%%%%%%%

%%%%%%%%%%% 
\section{The sample objects} 
\label{sec:sample} 
%%%%%%%%%%%

For ensuring the least possible contribution of N$^{2+}$ to the total N abundance of the nebulae, we have revised the literature to set up a sample of {\hii} regions showing a very low ionisation degree. In particular, we selected 13 objects where the spectroscopical observations available indicate a {\foiii}~5007/H$\beta$ =  0 or, at maximum, lower than 0.05. This sample comes from objects published in \citet{fichsilkey91}, \citet{hunter92}, \citet{glushkov95}, \citet{vilchezesteban96} and \citet{caplanetal00}. In general, as we will see in Section~\ref{sec:lines}, our observations confirm the absence of the {\foiii}~5007 \AA\ line or a {\foiii}~5007/H$\beta$  ratio lower than 0.05 in all the objects except in the cases of Sh~2-90 and Sh~2-152. \citet{hunter92} did not report the detection of {\foiii}~5007 \AA\ in none of the apertures she observed in Sh~2-90. However, we find {\foiii}~5007/H$\beta$  $\sim$ 0.75 in the brightest area of the nebula, that coincides with the position of her aperture no. 4. Sh~2-90 has a very large reddening coefficient --the largest of the whole sample-- and perhaps \citet{hunter92} did not detect the {\foiii}~5007 \AA\ line in her much  shallower spectra taken with the KPNO No. 2 0.9 m telescope. In the case of Sh~2-152 the explanation is simpler. In principle, the low-ionisation {\hii} region selected was Sh~2-153, which is a  faint and very diffuse nebula that was observed by \citet{caplanetal00} but lies very close to the much brighter Sh~2-152. Unfortunately, we commit the mistake of preparing the finding chart for Sh~2-152 instead of Sh~2-153 and the spectra was taken for the wrong object.

Following the same procedure as in \citet{estebanetal17}, the galactocentric distances, $R_{\rm G}$, assumed for the sample objects have been estimated as the mean values of kinematic and stellar distances given in different published references. We have associated an uncertainty for each distance, which corresponds to the standard deviation of the values considered for calculating the mean. We have assumed the Sun located at $R_{\rm G}$ = 8.0 kpc \citep{reid93} to make the calculations. To compute the mean distance and standard deviation of $R_{\rm G}$ for each object, we have used the kinematic distances determined by \citet{quirezaetal06}, \citet{balseretal11} and \citet{andersonetal15}, the stellar ones calculated by \citet{fosterbrunt15} and the kinematic and stellar distances calculated or compiled by \citet{russeil03} and \citet{caplanetal00}. In addition, for some objects, we have also considered distance determinations based on the comparison of the color-magnitude diagram of the brightest stars of their associated clusters with stellar evolutionary models. These have been the cases of Sh 2-237 \citep{pandeyetal13, limetal15} and Sh 2-266 \citep{mehneretal16}. 
Since kinematical distances have typical errors of the order of 20-25{\%} \citep{quirezaetal06, balseretal11}, our finally assumed uncertainties are of the order of 5-10\%. This is because the different distance determinations used for each object -- typically between 3 and 4 -- show a dispersion smaller than the individual errors.
The forthcoming second data release of Gaia mission will probably supersede these $R_{\rm G}$ determinations and reduce significantly their uncertainties. However, we do not expect significant changes in our general results -- except perhaps for some particular objects -- given the relatively low dispersion of the different $R_{\rm G}$ values given by the sources for each object.

Among other additional data that will be described in Section~\ref{sec:obs}, Table~\ref{tab:journal} gives the names and adopted $R_{\rm G}$ of the 13 {\hii} regions of our sample. This sample includes objects located from 5.7 to 16.1 kpc from the Galactic centre, covering a substantial fraction of the Galactic disc .

%%%%%%%%%%% 
\section{Observations and Data Reduction} 
\label{sec:obs} 
%%%%%%%%%%%

%%%%%%%%%%%%%%% 
\begin{figure*} 
\centering 
\includegraphics[scale=0.15]{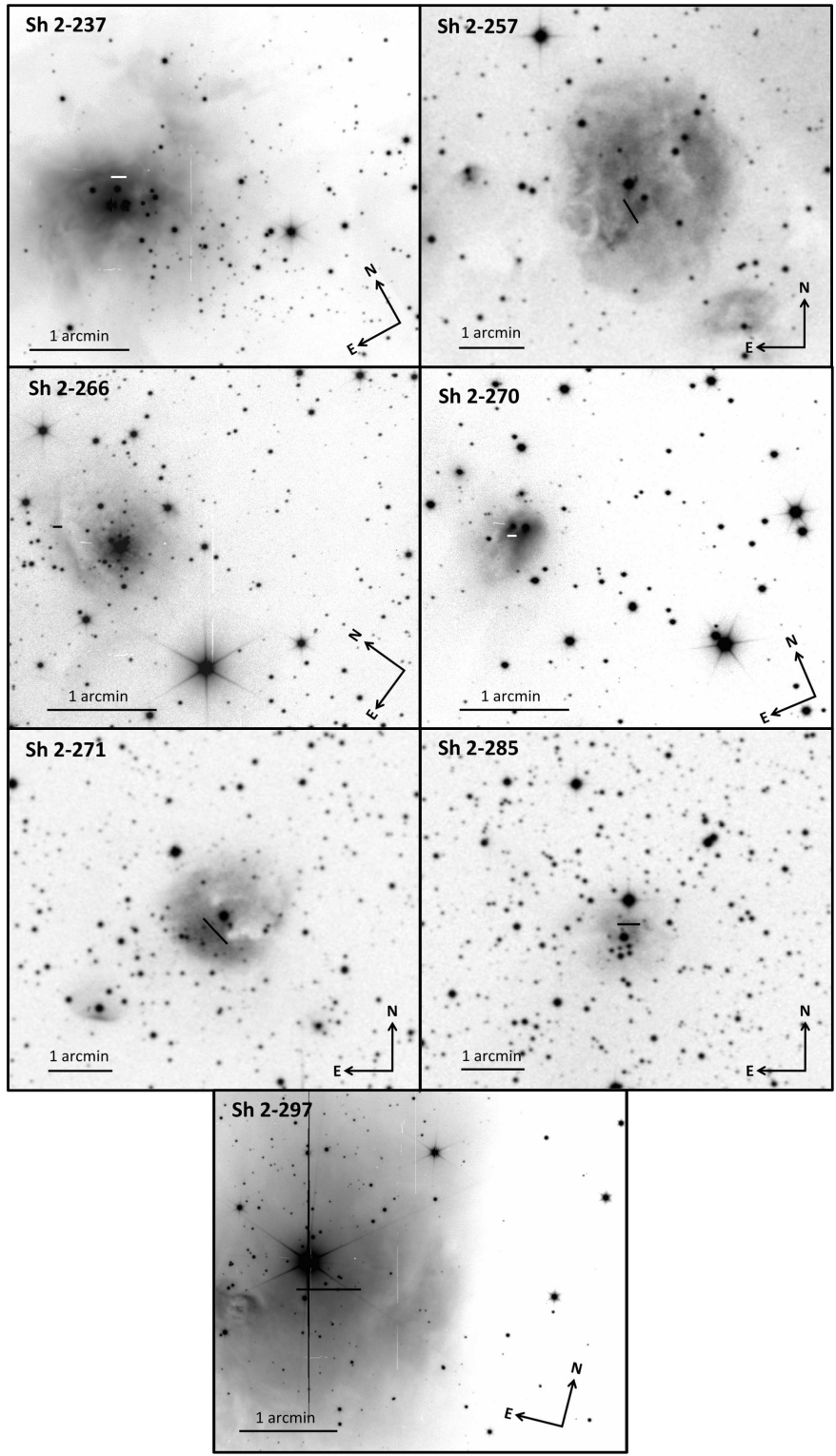} 
\caption{Finding charts of Sh~2-237, Sh~2-257, Sh~2-266, Sh~2-270, Sh~2-271, Sh~2-285 and Sh~2-297 indicating the position and length of the areas extracted for spectroscopical analysis (black/white lines). The charts of Sh~2-237, Sh~2-266, Sh~2-270 and Sh~2-297 are acquisition images taken with a {\it g} filter at the GTC. The charts of Sh~2-257, Sh~2-271 and Sh~2-285 correspond to red images of the Second Digitized Sky Survey (DSS-2) obtained from the Aladin interactive sky atlas \citep{bonnareletal00}.} \label{fig:slits_2}
 \end{figure*} 
 %%%%%%%%%%%%%%%

The observations were performed with the 10.4 m Gran Telescopio Canarias (GTC) at the Observatorio del Roque de los Muchachos (La Palma, Spain). They were carried out in 13 one-hour observing blocks distributed in several nights between September and December 2016. The spectra were taken with OSIRIS (Optical System for Imaging and low-Intermediate-Resolution Integrated Spectroscopy) spectrograph \citep{cepaetal00, cepaetal03}, which consists of a mosaic of two Marconi CCD42-82 CCDs each with 2048 $\times$ 4096 pixels and a 74 pixel gap between them. Each pixel has a physical size of 15$\mu$m. We used a binning 2 $\times$ 2 for our observations, giving a plate scale of 0.254 arcsec. OSIRIS was used in long-slit mode, centering the objects in CCD2. 
The slit length was 7.4 arcmin and its width was set at 0.8 arcsec. We used the R1000B grism to cover the whole optical spectral range, from 3600 to 7750 \AA\, achieving an effective spectral 
resolution of 6.52 \AA. The total integration time of the spectra was 2646 s for all objects, divided in three consecutive exposures of 882 s each.  This was done in order to accommodate the observations of each object in a standard one-hour observing block of the service mode. An additional 60 s exposure was taken in order to avoid problems with the 
eventual  saturation of the brightest lines. In figures~\ref{fig:slits_1} and \ref{fig:slits_2} we indicate the position and length of the apertures extracted for each object. In 
general, the apertures were chosen to cover the brightest zones of the nebula but also taking care that their extension optimizes the signal-to-noise ratio of the intensity of {\fnii} 5755 \AA\ line, which is necessary for the determination of the electron temperature (see Section~\ref{sec:conditions}). Sh~2-90 and  Sh~2-266 show some variation of the ionisation conditions -- $I$(\foiii)/$I$({\foii} line intensity ratios -- along the slit. In these two objects, the aperture coincides with the area showing the lowest ionisation combined with the best possible measurement of the {\fnii} 5755 \AA\ line. Considering that (a) normal {\hii} regions -- i. e. those do not hosting evolved massive Wolf-Rayet stars, as it is the case of the sample objects -- are expected to be composed by material with the original chemical composition of the cloud that formed the current ionising stars, and (b) the absence of observations indicating otherwise in the literature, we can assume that the chemical composition of any point of the {\hii} region is representative of the whole nebula.

Table~\ref{tab:journal} shows the coordinates of the centre of the extracted aperture, the Galactocentric distance of the object, position angle (PA), the aperture size of the extraction, the mean airmass during the total acquisition time of the spectrum and the mean H$\alpha$ surface brightness we measure in the areas extracted for spectroscopical analysis. The airmasses of the observations go from 1.05 to 1.56, so the zenith distance has never been larger than 50$^\circ$. The atmospheric differential refraction is not a problem for these observations because the objects are very extended -- several tens of arcseconds or even arcminutes -- and we do not expect that the conditions of the gas change in spatial scales of few arcseconds. 

The OSIRIS spectra were reduced using {\sc iraf}\footnote{{\sc iraf}, the Image Reduction and Analysis 
Facility, is distributed by the National Optical Astronomy Observatory, 
which is operated by the Association of Universities for Research 
in Astronomy under cooperative agreement with the National Science 
Foundation.} v2.16. Data reduction followed the standard procedure for long-slit 
spectra. The spectrograms were wavelength calibrated with Hg-Ar, Ne and Xe lamps. The absolute flux calibration was achieved by observations of the standard stars Ross 640, G191$-$B2B, Feige 110 and Hiltner 600. Particular care was taken in background subtraction because the sky 
background emission is not completely homogeneous along the GTC OSIRIS long slit \citep{fangetal15}. Since our targets are all 
extended sources, we take strips of the spectrograms free of nebular emission and multiply them by a factor changing slightly around one until the subtraction of the sky emission features are satisfactorily removed. As it has been said before, the spectrogram extends along two CCDs and, in some objects, the nebular emission fills completely the CCD  where the slit center is located (CCD2). The sky emission strips were taken from CCD1 in these cases (Sh~2-219, Sh~2-235, Sh~2-237 and Sh~2-257).

%%%%%%%%%%% 
\section{Line intensity measurements} 
\label{sec:lines} 
%%%%%%%%%%%

\begin{figure*} 
\centering 
\includegraphics[scale=0.19]{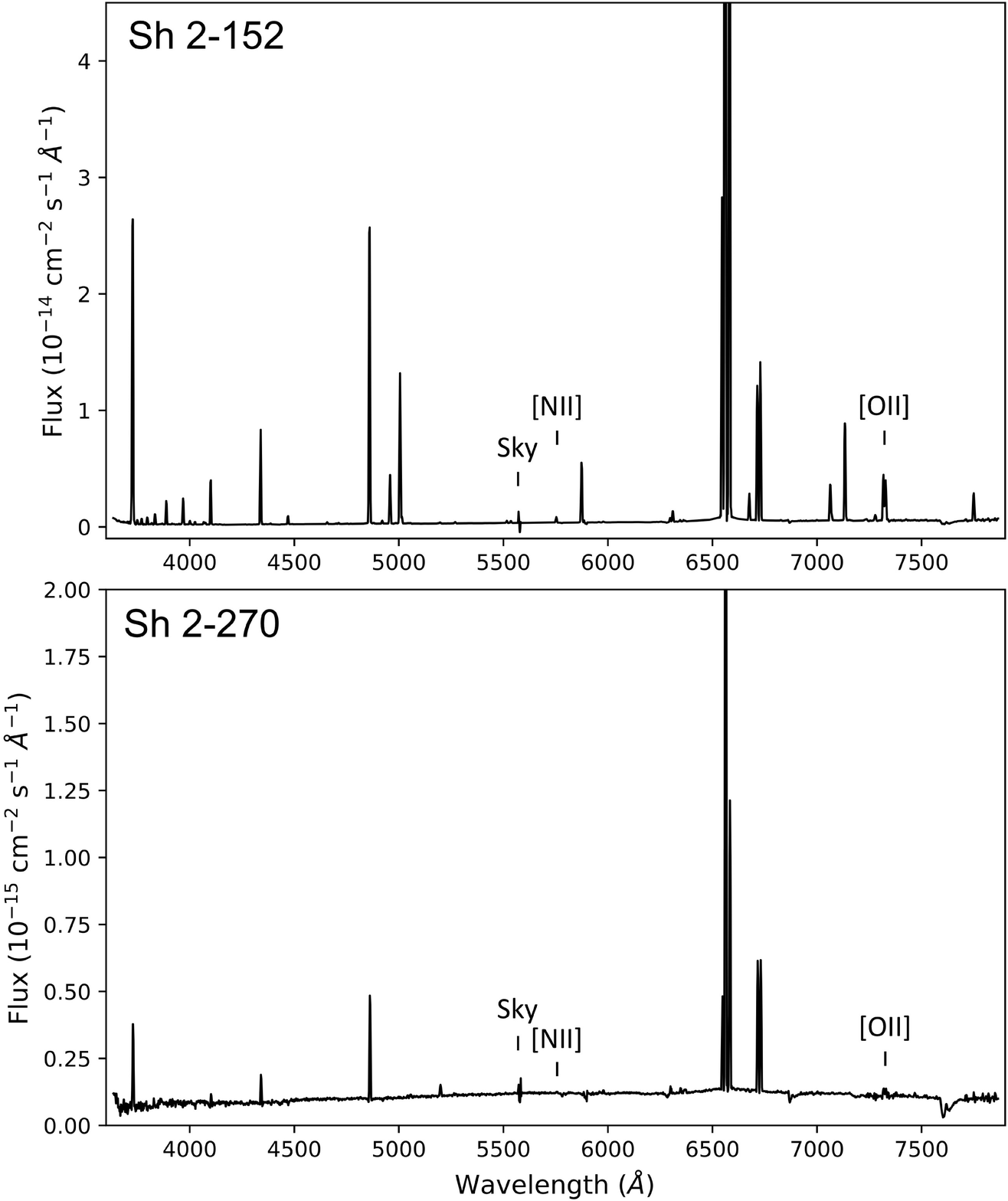} 
\caption{Flux-calibrated GTC OSIRIS spectra of Sh~2-152 (upper panel) and Sh~2-270 (bottom panel). These two examples correspond to the spectra with the highest and lowest signal-to-noise ratio of the sample, respectively. The {\foi} 5577 \AA\ sky emission feature and the 
{\fnii} 5755 \AA\ and {\foii} 7319, 7330 \AA\ lines are indicated.} 
\label{fig:spectra}
 \end{figure*}

Line fluxes of the spectra of the {\hii} regions included in Table~\ref{tab:journal} were measured with the {\sc splot} routine of {\sc iraf} by integrating all the flux in the line between two given limits and over the average local continuum. All line fluxes of a given spectrum have been normalized to H$\beta$ = 100.0. In the case of line blending, we applied a double or multiple Gaussian profile fit procedure using the {\sc splot} routine of {\sc iraf} to measure the individual line intensities. The identification of the lines were made following our previous results on spectroscopy of bright Galactic \hii\ regions \citep[see][and references therein]{garciarojasesteban07}. The number of lines detected and identified in the objects depends mainly on their surface brightness and varies from 66 in Sh~2-152 and 17 in Sh~2-270. In Figure~\ref{fig:spectra} we show the spectra of these two nebulae.

The reddening coefficient, $c$({\hb}), was determined from the comparison of the observed flux ratio of the brightest Balmer lines -- H$\alpha$, H$\gamma$ and H$\delta$ -- with respect to H$\beta$ and the theoretical line ratios computed by \citet{storeyhummer95} for the physical conditions derived for the nebulae -- see Section~\ref{sec:conditions} -- and following an iterative process. We have used the reddening function, $f(\lambda)$, normalized to H$\beta$ derived by \citet{cardellietal89} and assuming $R_V$ = 3.1.

%%%%%%%%%%%%%%% 
 %%%%%%%%%%%%%%%

In tables~\ref{tab:lines_1}, \ref{tab:lines_2} and \ref{tab:lines_3} of Appendix \ref{appex:1} we include the list of line identifications -- first 3 columns,  the reddening function, $f(\lambda)$ -- fourth column -- and dereddened flux line ratios with respect to H$\beta$ -- rest of columns. The quoted line intensity errors include the estimated flux calibration error ($\sim$ 2 percent), uncertainties in line flux measurement and error propagation in the reddening coefficient. Colons indicate line intensity errors of the order or greater than 40\%. The 2 last rows of each table include the reddening coefficient and the observed --uncorrected for reddening-- integrated H$\beta$ flux, $F$(H$\beta$), of the extracted aperture for each object.

%%%%%%%%%%% 
\section{Physical Conditions and chemical abundances} 
\label{sec:results} 
%%%%%%%%%%%

For the 13 objects of the sample, we have determined the physical conditions -- electron temperature, {\elect}, and density, {\elecd} -- and the ionic abundances making use of the version 1.0.26 of {\sc pyneb} \citep{Luridianaetal15} in combination with the atomic data listed in Table~\ref{tab:atomic} and the line-intensity ratios given in tables~\ref{tab:lines_1}, \ref{tab:lines_2} and \ref{tab:lines_3}.

%%%%%%%%%%%%% 
\begin{table*} 
\centering 
\caption{Atomic dataset used for collisionally excited lines.} 
\label{tab:atomic} 
\begin{tabular}{lcc} 
\hline 
& Transition probabilities &  \\ Ion & and energy levels & Collisional strengths \\ 
\hline 
N$^+$ & \citet{froesefischertachiev04} & \citet{tayal11} \\ 
O$^+$ & \citet{froesefischertachiev04} & \citet{kisieliusetal09} \\ 
O$^{2+}$ &  \citet{froesefischertachiev04, storeyzeippen00} & \citet{storeyetal14} \\ 
Ne$^{2+}$ & \citet{galavisetal97} & \citet{mclaughlinbell00} \\ 
S$^+$ & \citet{podobedovaetal09} & \citet{tayalzatsarinny10} \\ 
S$^{2+}$ &  \citet{podobedovaetal09} & \citet{tayalgupta99} \\ 
Cl$^{2+}$ & \citet{mendoza83} & \citet{butlerzeippen89} \\ 
Ar$^{2+}$ & \citet{mendoza83, kaufmansugar86} & \citet{galavisetal95} \\ 
Fe$^{2+}$ &  \citet{quinet96, johanssonetal00} & \citet{zhang96} \\ 
\hline 
\end{tabular} 
\end{table*} 
%%%%%%%%%%%%%

%%%%%%%%%%% 
\subsection{Physical Conditions } 
\label{sec:conditions} 
%%%%%%%%%%%

We have derived  {\elecd} using the density-sensitive emission line ratios of {\fsii}~6717/673 for all the objects but also {\fcliii}~5518/5538 in the cases of Sh~2-152 and 
Sh~2-90. We have assumed {\elecd}({\fsii}) as representative for all the objects and it is always below 1000 {\cmc}. The density derived from {\fcliii} lines has not been 
considered because its  much larger errors. We have determined {\elect} using {\fnii}~5755/(6548+6584) for all objects and {\foii}~(7319+7330)/(3726+3729) and {\fsii}~(4069+4076)/(6717+6731) for some of them. The quality of our measurements of the {\fnii} 5755 \AA\ line can be noted in figures~\ref{fig:5755_1} and \ref{fig:5755_2} of 
Appendix \ref{appex:2}. We have not corrected {\elect}({\foii}) and {\elect}({\fnii}) from the contribution to the intensity of {\foii} 7319, 7330 \AA\ and {\fnii} 5755 \AA\ lines due 
to recombination from O$^{2+}$ and N$^{2+}$ ions because it is very small or negligible in our sample objects. Most of the nebulae show an extremely low ionisation 
degree and the recombination contribution is clearly negligible. However, Sh~2-90 and Sh~2-152 show higher ionisation degrees, but the recombination contributions are also 
small in these cases. For Sh~2-90, the contribution is about 2\% and 0.5\% for the intensity of {\foii} and {\fnii} lines, respectively. For Sh~2-152, these percentages are even 
smaller, 0.6\% and 0.2\% respectively. In both cases the corrections are considerably smaller than the line intensity errors. The temperature-sensitive {\foiii}~4363 \AA\ line is 
only detected in Sh~2-152, however the uncertainty of this  measurement is too large to obtain reliable results. For all the objects, we estimate \elect({\foiii}) using \elect({\fnii}) 
and the empirical relation between both temperatures given in equation 3 of \citet{estebanetal09}. The physical conditions of the {\hii} regions are presented in 
Table~\ref{cond_abun}.

The large aperture of GTC and the high efficiency of OSIRIS in the red part of the optical spectrum, has permitted to measure the temperature-sensitive auroral {\fnii} 5755 \AA\ line in all the objects. This is the first time that {\elect} is determined from the intensity of CELs for the sample objects except in the case of Sh~2-266. For this nebula, \citet{vilchezesteban96} determined {\elect}({\foii}) for their inner and outer parts. Their slit position B is nearer to our aperture. Most of the sample objects have previous {\elect} determinations from radio recombination lines and continuum observations. All the {\elect} values available in the literature are shown in Table~\ref{tab:Te}. In general, our optical determinations are rather consistent with the radio ones considering the errors.

%%%%%%%%%%%%% 
\begin{table*} 
\centering 
\caption{Physical conditions and abundances for the sample objects.} 
\label{cond_abun} 
\begin{tabular}{l c c c c c c c} 
\hline 
& Sh2-61 & Sh2-152 & Sh2-235 & Sh2-271& Sh2-297 & Sh~2-90 & Sh~2-219\\ 
\hline 
\multicolumn{8}{c}{Physical conditions$^{\rm a}$} \\ 
\\ 
{\elecd}({\fsii}) & 1100 $\pm$ 170 &  750 $\pm$ 80 & 120 $\pm$ 40 & 90 $\pm$ 30 & 100 $\pm$ 40 & 150 $\pm$ 40 & $<$100\\ 
{\elecd}({\fcliii}) & $-$ & 870 $\pm$ 330 & $-$ & $-$ & $-$ & 3530 $\pm$ 2560 & $-$ \\ 
{\elect}({\fnii}) & 7580  $\pm$  330 & 8290  $\pm$  90 & 8130 $\pm$  190 & 8710 $\pm$  240 & 7830 $\pm$  200 & 8240 $\pm$  230 & 8840 $\pm$  320 \\ 
{\elect}({\foii}) & 9100  $\pm$  320 &  9130  $\pm$  260 & 9360 $\pm$  380 & 10260 $\pm$  490 & 9620 $\pm$  290  & 10780 $\pm$ 670 & 10840 $\pm$ 1030 \\ 
{\elect}({\fsii}) &  8670 $\pm$  1310 &  11680  $\pm$  860 & 11910 $\pm$  1520 & $-$ & $-$ & $-$ & $-$\\ 
{\elect}({\foiii})$^{\rm b}$ & 6380  $\pm$  330 &  7380  $\pm$  90 & 7150  $\pm$  190 & 7970  $\pm$  240 & 6730  $\pm$  200 & 7310 $\pm$  230 & 8160 $\pm$  320 \\ 
\\ 
\multicolumn{8}{c}{Ionic abundances$^{\rm c}$, O/H$^{\rm d}$, N/H$^{\rm d}$ and N/O$^{\rm e}$ ratios} \\ 
\\ 
He$^+$ & $-$ & 10.80 $\pm$ 0.01 & 10.59 $\pm$ 0.01 & 10.22 $\pm$ 0.03 & 10.17 $\pm$ 0.04 & 10.92 $\pm$ 0.01 & 10.35 $\pm$ 0.04 \\ 
N$^+$ & 7.87 $\pm$ 0.05 & 7.55 $\pm$ 0.02 & 7.55 $\pm$ 0.03 & 7.49 $\pm$ 0.03 & 7.69 $\pm$ 0.03 & 7.73 $\pm$ 0.03 & 7.41 $\pm$ 0.04 \\ 
O$^+$ & 8.39 $\pm$ 0.13 & 8.43 $\pm$ 0.03 & 8.39 $\pm$ 0.08 & 8.26 $\pm$ 0.07 & 8.45 $\pm$ 0.07 & 8.33 $\pm$ 0.09 & 8.22 $\pm$ 0.10 \\ 
O$^{2+}$ & 5.81: &  7.68 $\pm$  0.02 & 6.88 $\pm$  0.05 & 5.91 $\pm$  0.11 & 6.69 $\pm$  0.10 & 7.95 $\pm$ 0.06 & 5.97 : \\ 
S$^+$ & 6.77 $\pm$  0.05 & 6.13 $\pm$  0.02 & 6.33 $\pm$  0.03 & 6.20 $\pm$  0.03 & 6.46 $\pm$  0.03 & 6.32 $\pm$ 0.03 & 6.27 $\pm$ 0.04 \\ 
S$^{2+}$ & $-$ & 7.11 $\pm$  0.05 & 6.90 $\pm$  0.11 & 6.59 $\pm$  0.13 & $-$ & $-$ & $-$ \\ 
Cl$^{2+}$ & $-$ & 5.16 $\pm$  0.03 & 4.95 $\pm$  0.09 & 4.83 $\pm$  0.11 & 4.97 $\pm$  0.13 & 5.33 $\pm$  0.10 & 4.92 :  \\ 
Ar$^{2+}$ & $-$ & 6.18 $\pm$  0.04 & 5.92 $\pm$  0.06 & 5.10 $\pm$  0.11 & 5.54 $\pm$  0.09 & 6.32 $\pm$  0.07 & 5.23 $\pm$  0.11 \\ 
Fe$^{2+}$ & 6.30 $\pm$  0.19 & 6.02 $\pm$  0.02 & 5.67 $\pm$  0.13 & 6.02 $\pm$  0.09 & 5.99 $\pm$  0.13 & $-$ & $-$ \\ 
\\ 
O & 8.39 $\pm$ 0.13 &  8.50 $\pm$  0.03 & 8.41 $\pm$ 0.08 & 8.26 $\pm$ 0.07 & 8.45 $\pm$ 0.07 & 8.48 $\pm$ 0.07 & 8.22 $\pm$ 0.10 \\ 
N & 7.87 $\pm$ 0.05 & 7.63$^{\rm f}$ $\pm$  0.05 & 7.56$^{\rm f}$ $\pm$ 0.05 & 7.49 $\pm$ 0.03 & 7.70$^{\rm f}$ $\pm$ 0.03 & 7.88$^{\rm f}$ $\pm$ 0.08 & 7.41 $\pm$ 0.04 \\ 
N/O & $-$0.52 $\pm$ 0.14 & $-$0.87 $\pm$ 0.04 & $-$0.85 $\pm$ 0.08 & $-$0.77 $\pm$ 0.08 & $-$0.76 $\pm$ 0.08 & $-$0.61 $\pm$ 0.10 & $-$0.81 $\pm$ 0.11 \\ 
\hline 
& Sh2-237 & Sh2-257 & Sh2-266 & Sh2-175 & Sh2-270 & Sh~2-285 & \\ 
\hline 
\multicolumn{7}{c}{Physical conditions$^{\rm a}$}  &  \\ 
\\ 
{\elecd}({\fsii}) & 390 $\pm$ 60 &  110 $\pm$ 60 & 310 $\pm$ 200 & $<$100 & 440 $\pm$ 120 & $<$100  & \\ 
{\elect}({\fnii}) & 8860  $\pm$  260 & 7970  $\pm$  260 & 8350 $\pm$  480 & 7240 $\pm$  270 & 9510 $\pm$  920 & 	8640 $\pm$  310  &  \\ 
{\elect}({\foii}) & 9600  $\pm$  390 &  8680  $\pm$  310 & 21800 $\pm$  3200 & $-$ & 8580 $\pm$  1000 & $-$  & \\ 
{\elect}({\fsii}) &  $-$ & $-$ &  $-$ & $-$ & $-$ & 7780 $\pm$  1030  & \\ 
{\elect}({\foiii})$^{\rm b}$ & 8180  $\pm$  260 &  6930  $\pm$  260 & 7460  $\pm$  480 & 5900  $\pm$  270 & 9100  $\pm$  920 & 7870 $\pm$  310  & \\ 
\\ 
\multicolumn{7}{c}{Ionic abundances$^{\rm c}$, O/H$^{\rm d}$, N/H$^{\rm d}$ and N/O$^{\rm e}$ ratios}  & \\ 
\\ 
He$^+$ & 9.67 $\pm$ 0.07 & 10.24 $\pm$ 0.02 & 10.32 $\pm$ 0.07 & $-$ & $-$ & $-$  & \\ 
N$^+$ & 7.51 $\pm$ 0.03 & 7.58 $\pm$ 0.04 & 7.50 $\pm$ 0.08 & 7.69 $\pm$ 0.06 & 7.22 $\pm$ 0.10 & 7.42 $\pm$ 0.04  & \\ 
O$^+$ & 8.30 $\pm$ 0.08 & 8.38 $\pm$ 0.09 & 8.17 $\pm$ 0.16 & 8.42 $\pm$ 0.14 & 8.04 $\pm$ 0.28 & 8.19 $\pm$ 0.09  & \\ 
O$^{2+}$ & 6.34 $\pm$  0.10 & 6.54 $\pm$  0.09 & 6.05 $\pm$  0.16 & $-$ & $-$ & $-$  & \\ 
S$^+$ & 6.41 $\pm$  0.03 & 6.51 $\pm$  0.04 & 6.59 $\pm$  0.08 & 6.69 $\pm$  0.06 & 6.27 $\pm$  0.10 & 6.47 $\pm$ 0.04  & \\ 
S$^{2+}$ & $-$ & 6.97 $\pm$  0.31 & $-$ & $-$ & $-$ & $-$  & \\ 
Cl$^{2+}$ & 4.95 $\pm$  0.22 & 4.87 $\pm$  0.14 & $-$ & $-$ & $-$ & $-$  & \\ 
Ar$^{2+}$ & 5.12 $\pm$  0.10 & 5.46 $\pm$  0.10 & $-$ &  $-$ & $-$ & $-$  & \\ 
Fe$^{2+}$ & 5.78 $\pm$  0.19 & 5.56 $\pm$  0.20 & 6.32 $\pm$  0.32 & $-$ & $-$ & $-$  & \\ 
\\ 
O & 8.31 $\pm$ 0.07 &  8.38 $\pm$  0.09 & 8.18 $\pm$ 0.16 & 8.42 $\pm$ 0.14 & 8.04 $\pm$ 0.28 & 8.19 $\pm$ 0.09  & \\ 
N & 7.51 $\pm$ 0.05 & 7.58 $\pm$  0.04 & 7.51 $\pm$ 0.08 & 7.69 $\pm$ 0.06 & 7.22 $\pm$ 0.10 & 7.42 $\pm$ 0.04  & \\ 
N/O & $-$0.80 $\pm$ 0.08 & $-$0.80 $\pm$ 0.10 & $-$0.67 $\pm$ 0.18 & $-$0.73 $\pm$ 0.15 & $-$0.82 $\pm$ 0.30 & $-$0.77 $\pm$ 0.10  & \\ 
\hline 
\end{tabular} 
\begin{description} 
\item[$^{\rm a}$] {\elecd} in {\cmc}; {\elect} in K. 
\item[$^{\rm b}$] Estimated from {\elect}({\fnii}) and equation 3 of \citet{estebanetal09}. 
\item[$^{\rm c}$] In units of 12+log(\ionic{X}{n+}/\ionic{H}{+}). 
\item[$^{\rm d}$] In units of 12+log(X/H). 
\item[$^{\rm e}$] In units of log(N/O). 
\item[$^{\rm f}$] N/H ratio determined assuming the ICF(N$^+$) scheme of \citet{peimbertcostero69}.
\end{description} 
\end{table*} 
%%%%%%%%%%%%%

%%%%%%%%%%% 
\subsection{Abundances} 
\label{sec:abundances} 
%%%%%%%%%%%

Using CELs, we have derived abundances of N$^+$, O$^+$ and S$^+$ for all the objects and O$^{2+}$, S$^{2+}$, Cl$^{2+}$, Ar$^{2+}$ and Fe$^{2+}$ for some of them. We also determined the abundances of He$^+$ for several of the observed objects from the intensity of recombination lines. We assumed a two zone scheme for cal\-cu\-la\-ting the ionic abundances. We used  \elect({\fnii}) for low ionisation potential ions (N$^+$, O$^+$, S$^+$ and Fe$^{2+}$) and \elect({\foiii})  for the high ionisation potential ones (He$^+$, O$^{2+}$, S$^{2+}$, Cl$^{2+}$ and Ar$^{2+}$). Although we determine {\elect}({\foii}) in many of the objects, we do not consider it because it is systematically larger -- except in one case -- than  \elect({\fnii}) perhaps due to contamination from sky emission, which is specially strong in the red part of the spectrum and always difficult to subtract. Positive residuals of these sky emission features are affecting the measurement of the {\foii} 7319, 7330 \AA\ lines producing an overestimation of {\elect}({\foii}). That is especially dramatic in the case of Sh~2-266, where {\elect}({\foii}) is too high for a normal {\hii} region and very discrepant with respect to the value reported by \citet{vilchezesteban96}. We assumed  {\elecd}({\fsii}) as representative for all ions. The computations were made with {\sc pyneb} and using the atomic data listed in Table~\ref{tab:atomic}. The He$^+$ abundance has been determined making use of {\sc pyneb} and the effective recombination coefficient computations by \citet{porteretal12, porteretal13} for {\hei} lines, where collisional and optical depth effects in the line intensities are included. The final adopted He$^{+}$ abundance is the weighted average of the ratios obtained from the individual brightest {\hei} lines measured in each object. Table~\ref{cond_abun} gives the ionic abundances -- as well as their uncertainties -- for all the observed objects.

%%%%%%%%%%%%% 
\begin{table*} 
\centering 
\caption{{\elect} determinations (in K).} 
\label{tab:Te} 
\begin{tabular}{lccc} 
\hline 
{\hii} region & {\elect}({\fnii})$^{\rm a}$ &  Other$^{\rm b}$ & Reference  \\ 
\hline 
Sh 2-90 & 8240 $\pm$ 230 & 7370 $\pm$ 70 & \citet{quirezaetal06} \\ 
&  & 7760 $\pm$ 90 & \citet{balseretal11} \\ 
Sh 2-152 & 8290 $\pm$ 90 & 8400 $\pm$ 800 & \citet{afflerbachetal97}  \\ 
&  & 10000 $\pm$ 1000 & \citet{rudolphetal06} \\ 
&  & 9400 $\pm$ 80 & \citet{balseretal11}  \\ 
Sh 2-219 & 8840 $\pm$ 320 & 8800 $\pm$ 900 & \citet{fichsilkey91} \\ 
Sh 2-235 & 8130 $\pm$ 190 & 8940 $\pm$ 170 & \citet{quirezaetal06} \\ 
&  & 8610 $\pm$ 140 & \citet{balseretal11}  \\ 
Sh 2-237 & 8860 $\pm$ 260 & 8830 $\pm$ 160 & \citet{balseretal11} \\ 
Sh 2-257 & 7970 $\pm$ 260 & 9100 $\pm$ 900 & \citet{shaveretal83} \\ 
&  & 8830 $\pm$ 110 & \citet{balseretal11}  \\ 
Sh 2-266 & 8350 $\pm$ 480 & 11100 $\pm$ 1100 & \citet{fichsilkey91} \\ 
&  & 10200 $\pm$ 200$^{\rm c}$ & \citet{vilchezesteban96}  \\ 
Sh 2-270 & 9510 $\pm$ 920 & 9500 $\pm$ 900 & \citet{fichsilkey91} \\ 
Sh 2-271 & 8710 $\pm$ 240 & 9100 $\pm$ 900 & \citet{fichsilkey91} \\ 
Sh 2-285 & 8640 $\pm$ 310 & 9700 $\pm$ 1000 & \citet{fichsilkey91} \\ 
Sh 2-297 & 7830 $\pm$ 200 & 7540 $\pm$ 160 & \citet{fichsilkey91} \\ 
\hline 
\end{tabular} 
\begin{description} 
\item[$^{\rm a}$] This work. 
\item[$^{\rm b}$] Based on radio recombination observations except indicated. 
\item[$^{\rm c}$] {\elect}({\foii}) determination. 
\end{description} 
\end{table*}

%%%%%%%%%%%%%

In {\hii} regions, O is commonly the only element for which no ionisation correction factor (hereafter ICF) is necessary to derive its total abundance. Therefore, the total O abundance is simply the sum of the O$^+$/H$^+$ and O$^{2+}$/H$^+$ ratios. In this paper, we select objects of very low ionisation degree for which the assumption N/H $\approx$ N$^+$/H$^+$ can be applied, i. e. no ICF is necessary to derive their total N abundance or, at least, the expected contribution of the N$^{2+}$/H$^+$ ratio is 
smaller than the uncertainty of the N$^+$ abundance. Exceptions of this situation are Sh~2-90 and Sh~2-152, for which non-negligible values of the  ICF have to be applied, see Sect.~\ref{sec:Ngradient} for details). Therefore, with the strategy outlined above we avoid the always difficult task of estimating uncertainties for abundances derived using ICFs substantially larger than one. With our data we can also estimate total abundances of He, S,  Cl, Ar and Fe for our objects. We plan to study the gradients of these elements in a future paper where we will perform a critical analysis of the best ICF sets including results of other Galactic {\hii} regions  presented in \citet{estebanetal17}. Table~\ref{cond_abun} gives the total N and O abundances as well as the N/O ratio of the observed objects.

%%%%%%%%%%% 
\section{The radial gradient of N} 
\label{sec:Ngradient} 
%%%%%%%%%%%

%%%%%%%%%%% 
\begin{figure*} 
\centering 
\includegraphics[scale=0.78]{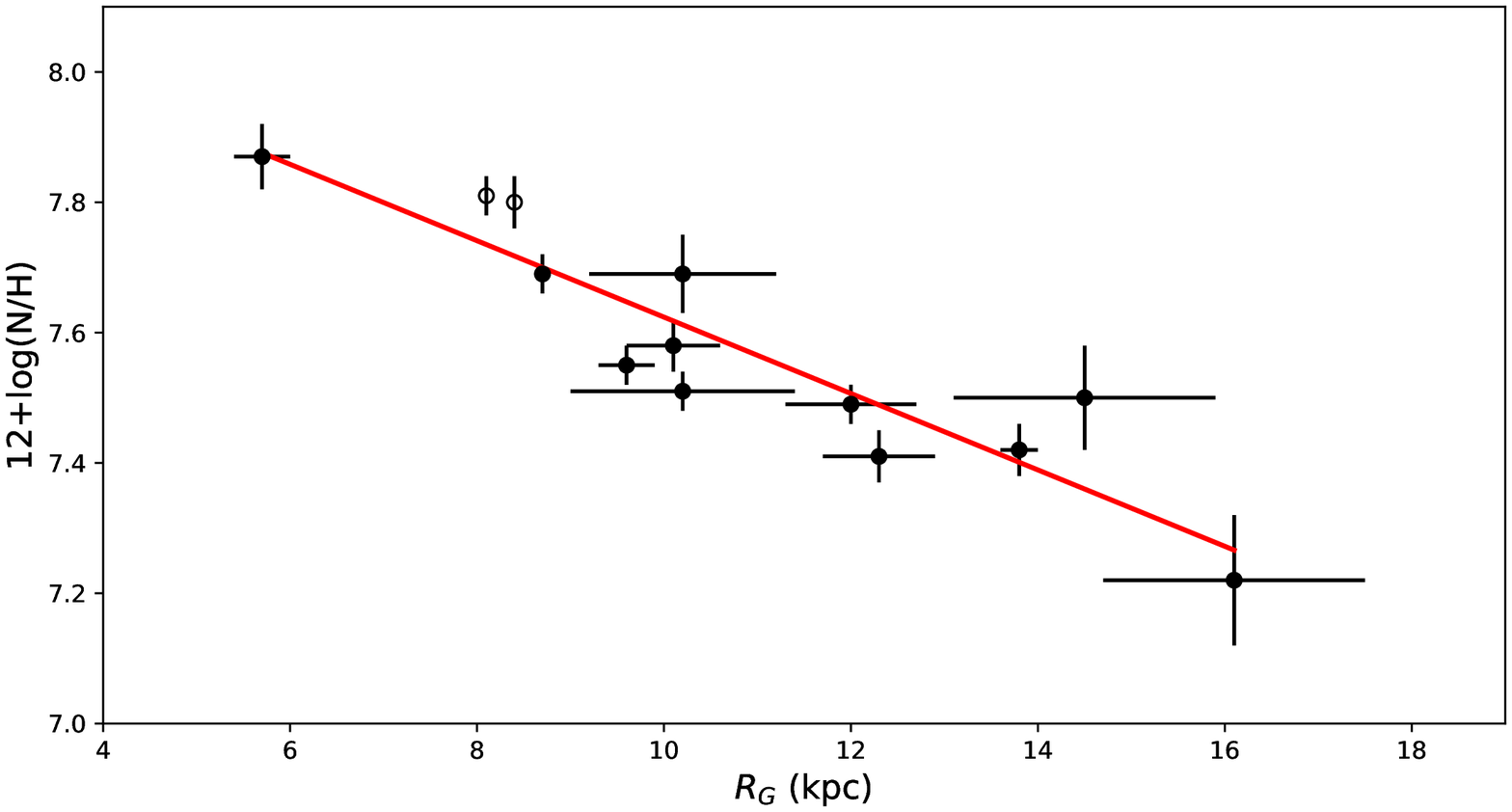} 
\caption{Radial distribution of the N abundance -- in units of 12+log(N/H) -- as a function of the Galactocentric distance, $R_{\rm G}$, for the sample of Galactic {\hii} regions studied in this paper except Sh~2-90 and Sh~2-152 -- full circles.  We can assume N/H $\approx$ N$^+$/H$^+$ for this group of very low ionisation nebulae. The empty circles correspond to data for other low ionisation objects taken from the literature:  IC~5146 and M~43 -- see text for references. The solid red line represents the least-squares fit to all the objects.} 
\label{fig:Ngrad} 
\end{figure*} 
%%%%%%%%%%%

As it can be seen in Table~\ref{cond_abun}, three of the objects show a null contribution of the O$^{2+}$/H$^+$ ratio, eight show a value of this quantity smaller than the estimated errors of the total O abundance and only Sh~2-90 and Sh~2-152 present a significant amount of O$^{2+}$/H$^+$. Considering the similarity of the ionisation potential of N$^+$ 
and O$^+$ (29.6 and 35.1 eV, respectively), we can assume that the amount of N$^{2+}$ is virtually negligible in all the objects except in Sh~2-90 and Sh~2-152. Of those very low-ionization nebula, only Sh~2-235 and Sh~2-297 show a N/H ratio somewhat higher than N$^+$/H$^+$, but this difference is almost marginal, about 0.01 dex, smaller than the uncertainties of the N$^+$ abundance\footnote{This small difference is similar whether we use the ICF(N$^+$) scheme of \citet{peimbertcostero69} or \citet{izotovetal06}. The application of the ICF(N$^+$) scheme of \citet{mathisrosa91} depends on the measurement of {\fsii} and {\fsiii} lines. {\fsiii} is not detected in Sh~2-297 and has a relatively high uncertainty in the case of Sh~2-235.}. Therefore, we can conclude that N/H $\approx$ N$^+$/H$^+$ within the uncertainties in all the very low-ionisation {\hii} regions. 

In Fig.~\ref{fig:Ngrad} we show the spatial distribution of the N abundances of the very low-ionisation {\hii} regions (all the objects of the sample except Sh~2-90 and Sh~2-152, full circles). We also include data for two other Galactic star-forming regions of the literature with direct determination of {\elect} that do not show {\foiii} emission (empty circles): IC~5146 and M~43. For IC~5146, we use the average abundances of apertures 2, 3 and 4 observed by \citet{garciarojasetal14} and recalculated by \citet{estebanetal17}. For M~43 we use the average abundances of apertures A4 and A5 observed by \citet{simondiazetal11} and corrected for contamination from the emission of the neighboring Orion Nebula.

We have performed a least-squares linear fit to the $R_{\rm G}$ of the very low-ionisation objects and their N abundance -- all the objects included in Table~\ref{cond_abun} except Sh~2-90 and Sh~2-152 -- covering values of $R_{\rm G}$ ranging from 5.7 to 16.1 kpc. We will name this subsample of nebulae as the ``non-ICF group''. The fit gives the following radial N  abundance gradient:

\begin{equation} \label{eq:1} 12 + \log(\mathrm{N/H}) = 8.21(\pm 0.09) - 0.059(\pm 0.009) R_\mathrm{G}. \end{equation}

The uncertainties of the slope and intercept of the linear fit are computed through Monte Carlo simulations. We generate 10$^4$ random values of $R_{\rm G}$ and the N abundance for each observational point assuming a Gaussian distribution with a sigma equal to the uncertainty of each quantity. We performed a least-squares linear fit to each of these 10$^4$ random distributions. It is important to remark that we have considered the uncertainty of distances in the fittings, which are not usually taken into account in most works. The uncertainties associated to the slope and intercept correspond to the standard deviation of the values of these two quantities obtained from the fits. The spatial distribution of the N abundances and the gradient are shown in Fig.~\ref{fig:Ngrad}. It is important to remark that the mean difference of the N abundance of the {\hii} regions represented in Fig.~\ref{fig:Ngrad} and the abundance given by the linear fit at their corresponding distance is $\pm$0.06 dex, of the order of the average uncertainty of the abundance determinations, which is about $\pm$0.05 dex. The maximum difference we find is of the order of $\pm$0.14 dex. This result indicates that the amount of N in the interstellar medium of the Galactic disc is fairly homogeneous and that any possible local inhomogeneity is not substantially larger than the observational uncertainties.

%%%%%%%%%%% 
\begin{figure} 
\centering 
\includegraphics[scale=0.37]{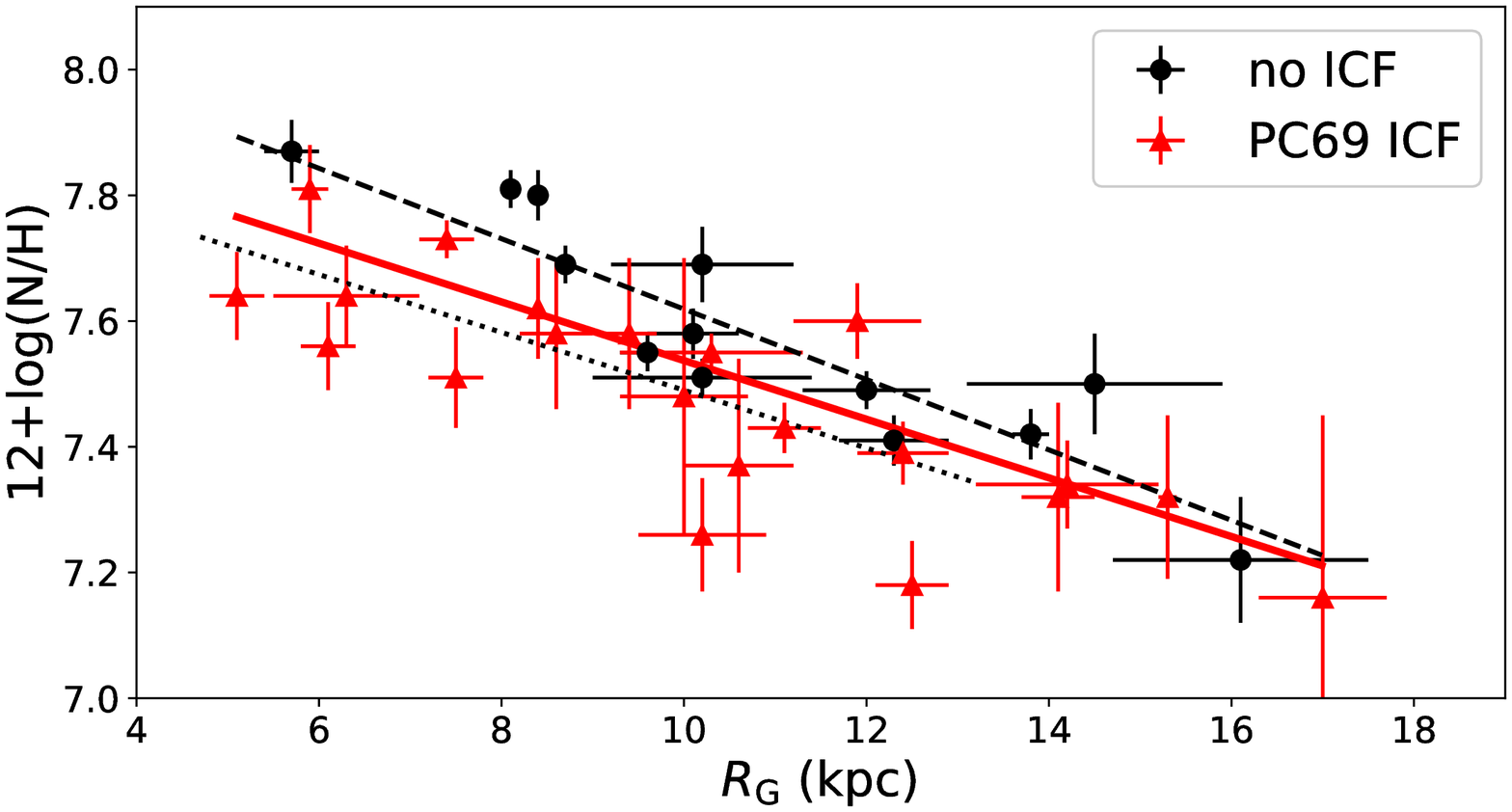} 
\includegraphics[scale=0.37]{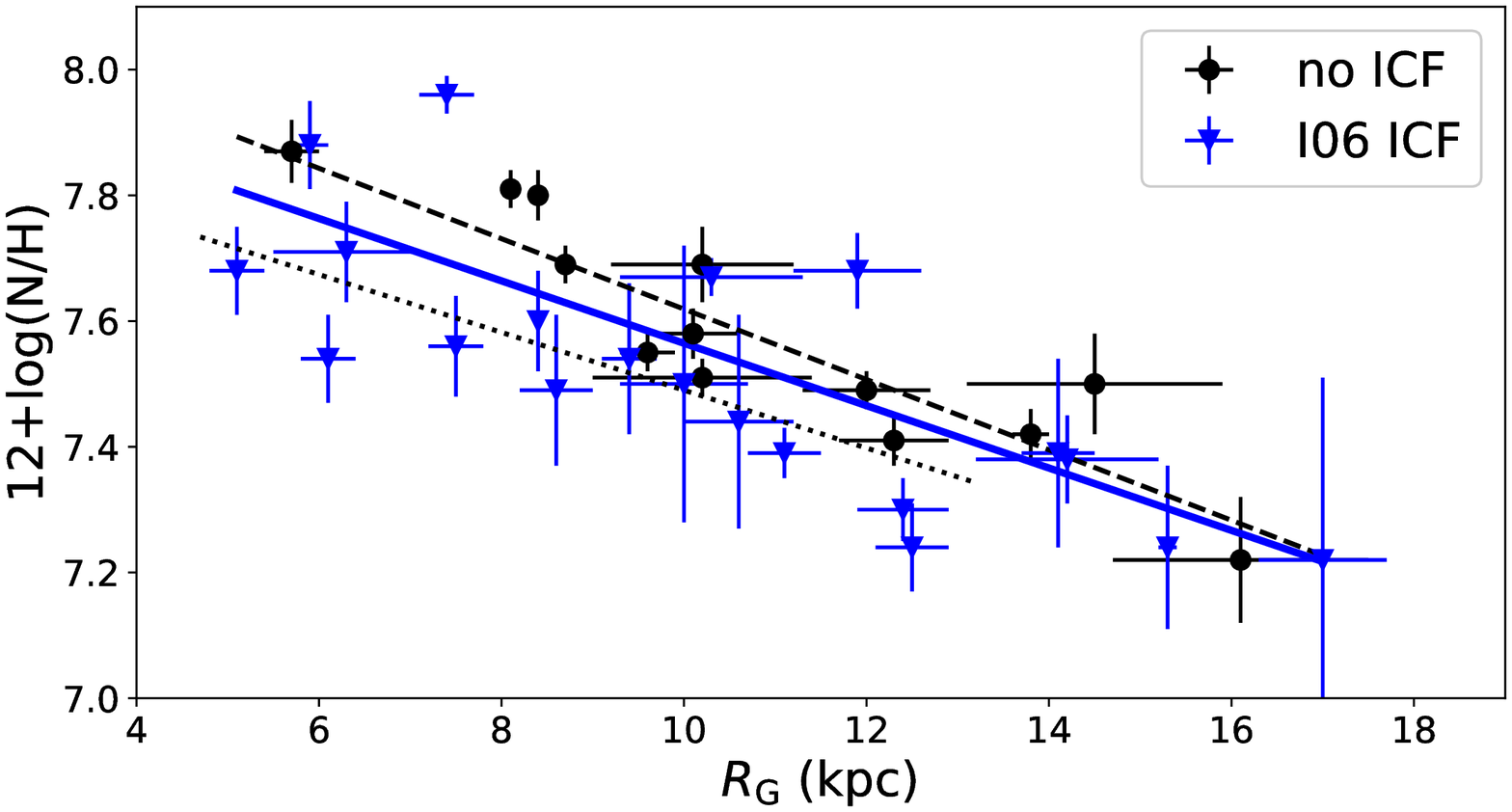} 
\includegraphics[scale=0.37]{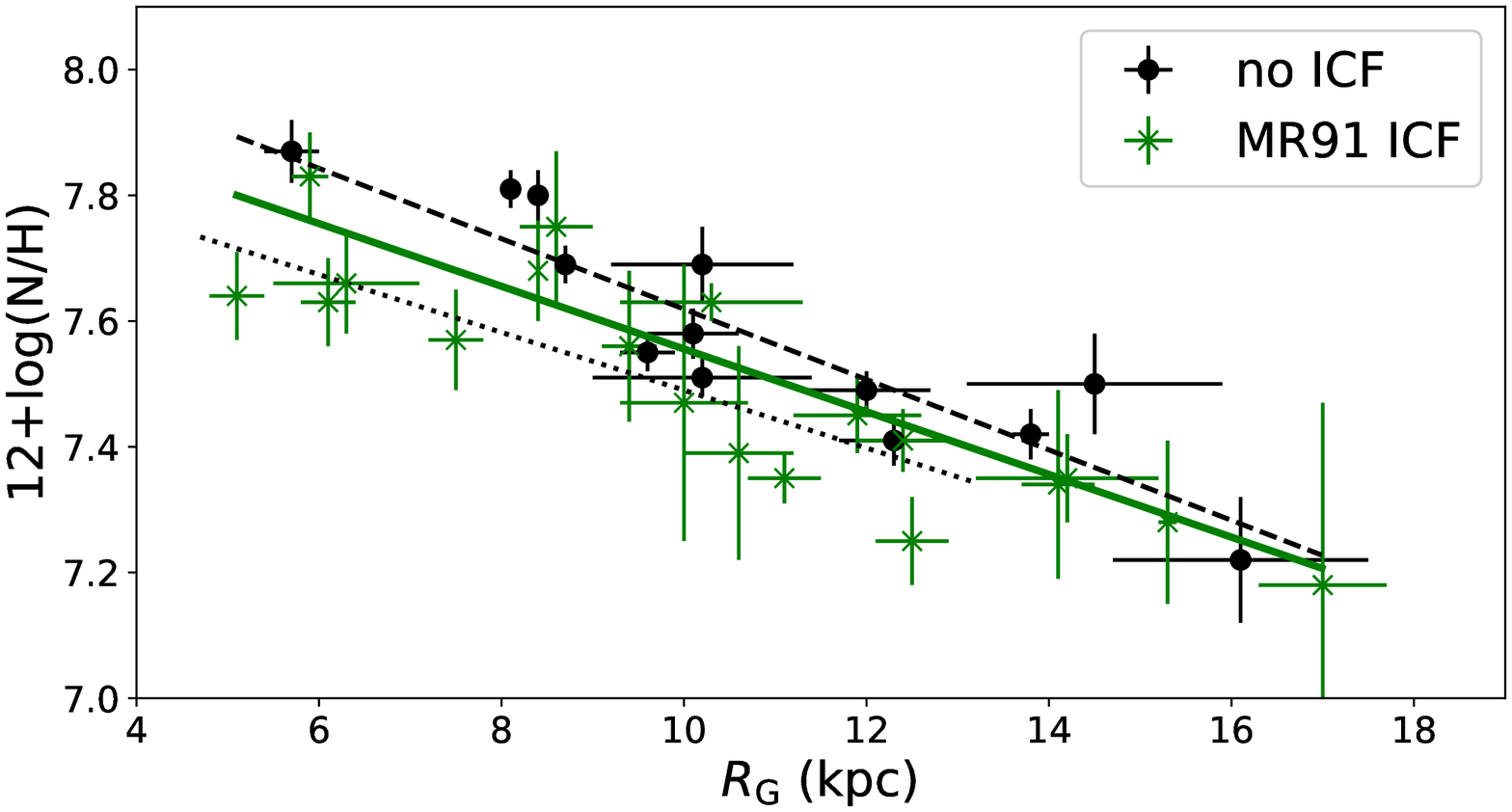} 
\caption{Radial distribution of the N abundance -- in units of 12+log(N/H) -- as a function of the Galactocentric distance, $R_{\rm G}$, for the Galactic {\hii} regions of the ICF and non-ICF groups. Black circles represent the objects of the non-ICF group -- those included in Fig.~\ref{fig:Ngrad} -- and the rest of coloured symbols indicate objects of the ICF group. We have assumed different ICF(N$^+$) schemes to derive N abundances of the objects of the ICF group. Upper panel: N/H ratio determined using the classical ICF(N$^+$) of \citet{peimbertcostero69}, PC69, red triangles. Middle panel: N/H ratio determined using the ICF(N$^+$) of \citet{izotovetal06}, I06, blue inverted triangles. Lower panel: N/H ratio determined using the ICF(N$^+$) of \citet{mathisrosa91}, MR91, green crosses. The coloured solid lines represent the least-squares fit to all the objects represented in each panel. The dashed black lines correspond to the least-squares fit to the objects of the non-ICF group. The dotted black lines represent the radial N gradient derived from B stars by \citet{dafloncunha04}.} 
\label{fig:Ngradall} 
\end{figure} 
%%%%%%%%%%%

We have made a further determination of the radial N gradient including the data of {\hii} regions of higher ionisation degree where the approximation N/H $\approx$ N$^+$/H$^+$ is not longer valid. The set of objects includes Sh~2-90, Sh~2-152  and the {\hii} regions of the whole sample used in \citet{estebanetal17}, for which ionic abundances have been determined using the same methodology and atomic data than in this paper. This subsample contains 21 nebulae and we will refer it as the ``ICF group''. The range of values of $R_{\rm G}$ covered by this set of objects goes from 5.1 to 17.0 kpc.

In order to derive the total N abundance of normal {\hii} regions  where some amount of N$^{2+}$ is present, an ionisation correction factor (ICF) has to be adopted. The ICF(N$^+$) is a multiplicative factor to transform N$^+$/H$^+$ ratios into N/H ones following the simple relation:

\begin{equation}\label{eq:2} \mathrm{N/H} = \mathrm{ICF}(\mathrm{N}^{+}) \times \mathrm{N}^{+}/\mathrm{H}^{+}. \end{equation}

In our study, we will use and intercompare three of the most widely used ICF(N$^+$) schemes in the literature. Firstly, we consider the scheme proposed by \citet{peimbertcostero69} based on the similarity of the ionisation potential of the ionic species of N with those of O. Secondly, the fitting expressions obtained by \citet{izotovetal06} from photoionisation models for extragalactic {\hii} regions. These last authors define different fittings depending on three metallicity ranges. In our case, we only use their high metallicity fit, valid for 12 + log(O/H) $\geq$ 8.2. Finally, the set of ICFs proposed by \citet{mathisrosa91} based on nebular photoionisation models that mainly use plane-parallel stellar atmospheres as ionisation sources. They consider two regimes of ``hot" and ``cold" Kurucz atmospheres.

The least-squares linear fit to the $R_{\rm G}$ and the N/H ratios including the objects of the non-ICF and ICF groups when using the ICF(N$^+$) scheme of \citet{peimbertcostero69}, gives the following radial N  abundance gradient: \begin{equation} \label{eq:3} 12 + \log(\mathrm{N/H}) = 8.00(\pm 0.07) - 0.047(\pm 0.008) R_\mathrm{G}. \end{equation}

This fit and the data points are represented in the upper panel of Fig.~\ref{fig:Ngradall}. In this case, the maximum difference in the N/H ratio between the data points and the line fit is 0.27 dex and the mean difference 0.09 dex. We can see that the fit given by Eq.~\ref{eq:3}  -- red continuous line -- is located between 0.02 and 0.13 dex below the line defined by the non-ICF group -- black dashed line -- and given by Eq.~\ref{eq:1}.

The least-squares linear fit to $R_{\rm G}$ and the N/H ratios calculated using the ICF(N$^+$) scheme of \citet{izotovetal06}, gives:

\begin{equation} \label{eq:4} 12 + \log(\mathrm{N/H}) = 8.06(\pm 0.07) - 0.050(\pm 0.008) R_\mathrm{G}. \end{equation}

The N abundance distribution and the fit obtained with this ICF(N$^+$) are represented in the middle panel of Fig.~\ref{fig:Ngradall}. In this case, the line fit is still below but closer to the one obtained for the non-ICF group -- between 0.01 and 0.09 dex --, indicating that the N abundances determined with the ICF(N$^+$) of  \citet{izotovetal06} are slightly more similar to those determined without ICF(N$^+$) at the same distance than the abundances obtained with the scheme of \citet{peimbertcostero69}. However, the maximum difference in the N/H ratio between the data points as well as the line fit and the mean difference are almost the same than in the previous case, 0.27 dex and 0.10 dex, respectively.

Finally, the least-squares linear fit to $R_{\rm G}$ and the N/H ratios calculated using the ICF(N$^+$) scheme of \citet{mathisrosa91}, gives:

\begin{equation} \label{eq:5} 12 + \log(\mathrm{N/H}) = 8.05(\pm 0.08) - 0.050(\pm 0.008) R_\mathrm{G}. \end{equation}

The N abundance distribution and the fit obtained with this ICF(N$^+$) are shown in the lower panel of Fig.~\ref{fig:Ngradall}. As we can see, the fits obtained with the ICF(N$^+$) schemes of \citet{izotovetal06} and \citet{mathisrosa91} are almost identical. However, the dispersion of the data points with respect to the line fit is smaller in the last case, the maximum difference is 0.18 dex and the mean difference 0.08 dex.

%%%%%%%%%%% 
\begin{figure} 
\centering \includegraphics[scale=0.37]{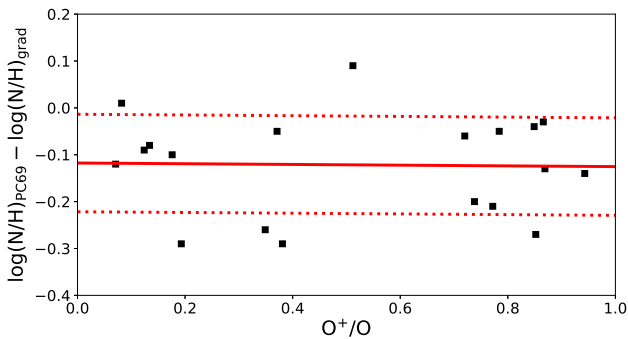} 
\caption{Difference between the N abundance we determine for the objects of the ICF group using the ICF(N$^+$) of \citet{peimbertcostero69} -- 12+log(N/H)$_{\rm PC69}$ -- and the N abundance corresponding to their $R_{\rm G}$ using the fit of Eq.~\ref{eq:1} -- 12+log(N/H)$_{\rm grad}$ -- as a function of their ionisation degree, O$^+$/O. The solid red line represents the least-squares fit to the points. The dotted lines indicate the standard deviation of the points around the line fit.} 
\label{fig:dif} 
\end{figure} 
%%%%%%%%%%%

As a general result of the different fits shown in Fig.~\ref{fig:Ngradall}, the N abundances of most {\hii} regions of the ICF group derived using any ICF(N$^+$) scheme are lower than the abundance given by the linear fit given in Eq.~\ref{eq:1} by factors as high as 0.3 dex but usually in the range between 0.05 and 0.10 dex. The dispersion of the data points around the line fit is typically of the order of 0.1 dex, larger than the typical abundance errors and the dispersion around the line fit obtained for the non-ICF group in Fig.~\ref{fig:Ngrad}. As expected, the use of an ICF(N$^+$) is a source of uncertainty for the determination of N abundances at a given distance.  Another general result is that the slopes of the linear fits obtained using any ICF(N$^+$) scheme are slightly flatter than the one obtained for the fit of the objects of the non-ICF group.

In Fig.~\ref{fig:Ngradall} we also include the radial N gradient obtained by \citet{dafloncunha04} from abundance determinations in young OB stars. The slope found by those authors is $-$0.042 $\pm$ 0.007 dex kpc$^{\rm -1}$, consistent with the slopes of the N gradients we find using ICF(N$^+$) schemes within the uncertainties. However, the stellar N abundances are systematically lower than the nebular ones, specially with respect to the abundances of the objects of the non-ICF group, which are between 0.1 and 0.2 dex higher than the stellar ones for the same value of $R_{\rm G}$. In principle, considering the relatively short lives of massive stars, we expect that OB stars and {\hii} regions located at the same $R_{\rm G}$ should show similar abundances. However, N can be enhanced in some OB stars due to mixing of CN-processed material into the atmospheric layers \citep[e. g.][]{przybillaetal10}. In this context, the fact that the mean N abundances of {\hii} regions are somewhat higher than those obtained for OB stars suggests that one or both kinds of objects are not giving the correct present-day N abundance of the Galactic disc. One would invoke the effect of dust depletion in the nebulae, but this would affect in the opposite direction lowering the nebular abundances. However, contrary to what happens with O, we do not expect that nebular abundances of N are significantly affected for dust depletion because this element is not a major constituent of dust \citep{gailsedlmayr86,jenkins14}.

Previous determinations of the radial abundance gradient of N from {\hii} regions give slopes stepper than the ones we have found in all our fits. \citet{shaveretal83} obtain $-$0.073 $\pm$ 0.013 dex kpc$^{\rm -1}$ and \citet{carigietal05} $-$0.085 $\pm$ 0.020 dex kpc$^{\rm -1}$, on the other hand, from FIR observations, \citet{rudolphetal06} 
obtain a slope of $-$0.085 $\pm$ 0.020 dex kpc$^{\rm -1}$. It is difficult to understand the reason because we obtain a flatter slope apart from the use of better observational data with respect to other authors. In the case of \citet{shaveretal83} the finding of a stepper slope 
is not a strange result because they also obtained a stepper radial O abundance gradient and used the ICF scheme of \citet{peimbertcostero69} to derive the N/H ratio. 
Both aspects would produce naturally a stepper N gradient. In the case of \citet{carigietal05} we can argue that they use a smaller sample distributed in a much narrower baseline of 
$R_{\rm G}$ (see discussion in Section~\ref{sec:NOgradient}), so the statistical significance of their gradient determination is lower than ours. Finally, as it will be discussed in 
Section~\ref{sec:NOgradient}, FIR studies use non cospatial {\elect} and line intensity determinations, and this can affect the precision of the abundances.

In Fig.~\ref{fig:dif} we plot the difference between the N abundance we determine for the objects of the ICF group using the standard ICF(N$^+$) of \citet{peimbertcostero69} and the N abundance we calculate at their $R_{\rm G}$ using Eq.~\ref{eq:1} -- the N gradient determined for the non-ICF group -- as a function of their ionisation degree, parameterized with O$^+$/O. We choose to represent the results obtained using this particular ICF(N$^+$) scheme because it gives the largest differences with respect to the results obtained for the non-ICF group. 
The slope of the linear least-squares fit to the points represented in Fig.~\ref{fig:dif} is $-$0.008, practically zero, indicating that the difference is independent of the ionisation degree of the gas. As it can be seen in Fig.~\ref{fig:dif}, the net effect of assuming our ``ICF-free'' gradient instead of the standard ICF(N$^+$) of \citet{peimbertcostero69} is an increase of the N/H ratio of about 0.12 dex, with a standard deviation of about 0.1 dex.

\begin{figure} 
\centering 
\includegraphics[scale=0.37]{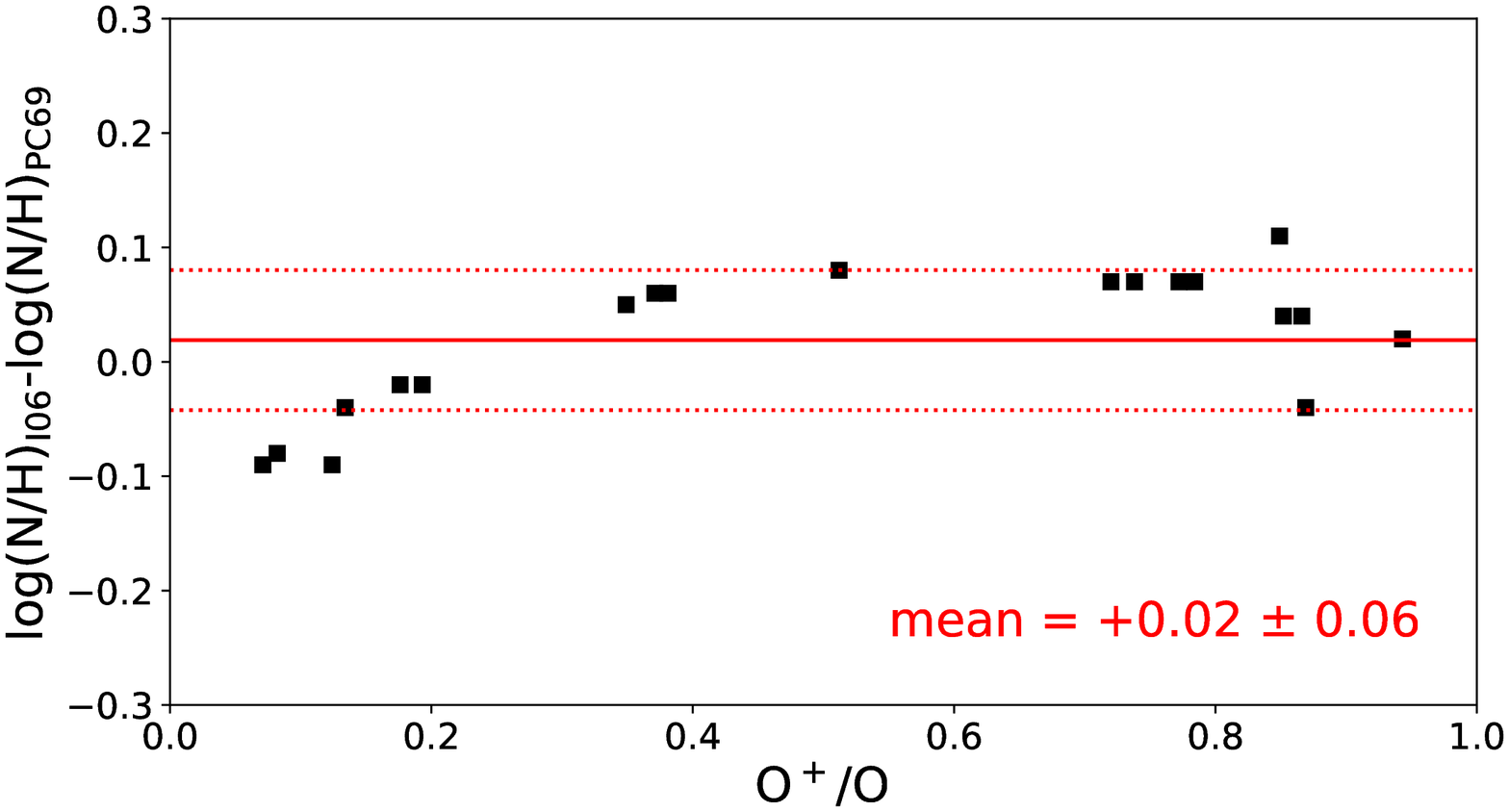} 
\includegraphics[scale=0.37]{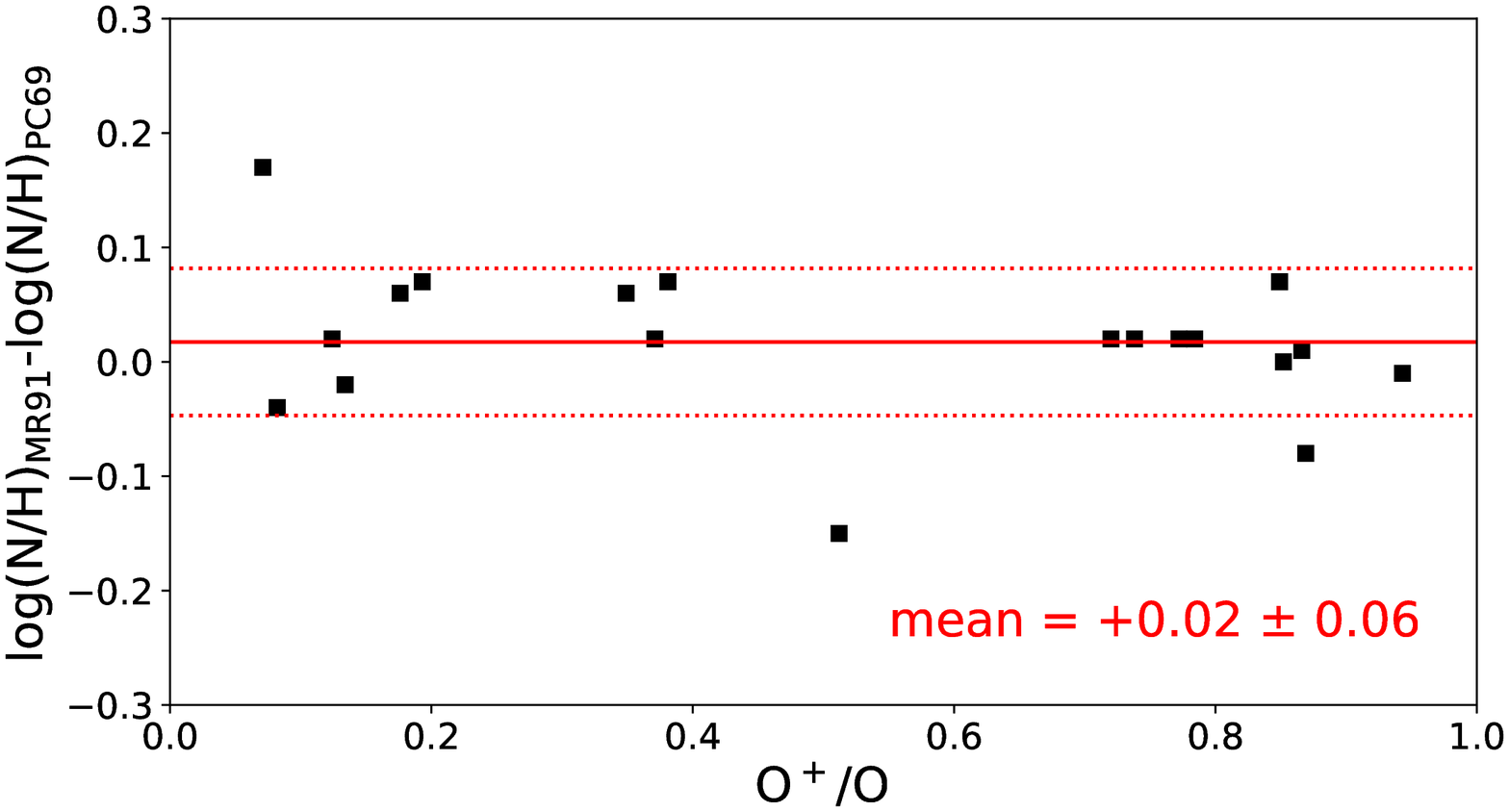} 
\includegraphics[scale=0.37]{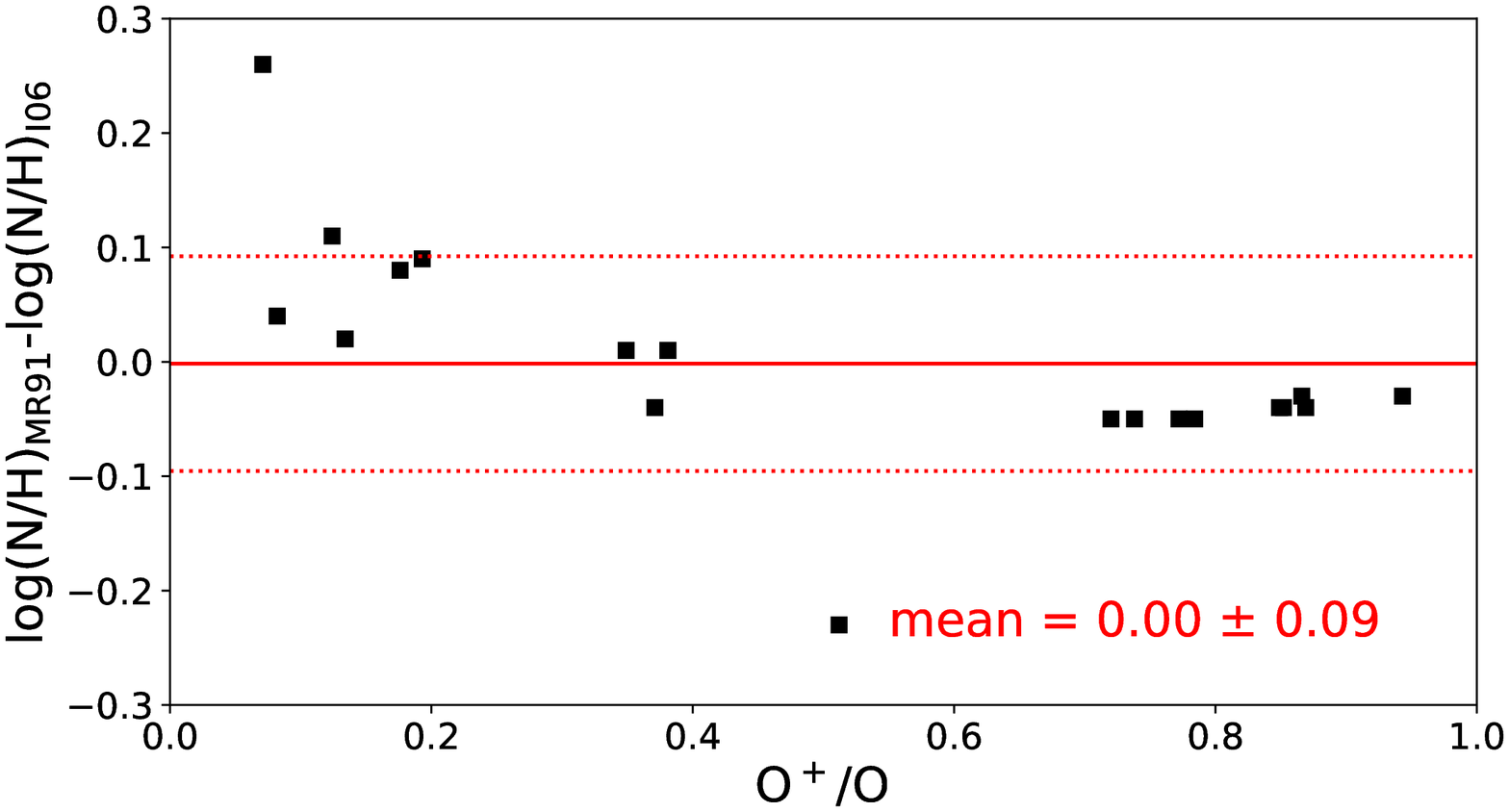} 
\caption{Difference between the N abundance determined for the objects of the ICF group using different ICF(N$^+$) schemes as a function of their ionisation degree, O$^{+}$/O. Upper panel: comparison of the N abundances obtained using the ICF(N$^+$) of \citet{izotovetal06} and \citet{peimbertcostero69}. Middle panel: \citet{mathisrosa91} and \citet{peimbertcostero69}. Lower panel: \citet{mathisrosa91} and \citet{izotovetal06}. The solid red line represents the mean value of the differences. The dotted lines indicate the standard deviation around the mean.} 
\label{fig:compCFs} 
\end{figure}

In Fig.~\ref{fig:compCFs} we intercompare the N/H ratios determined for the objects of our ICF group using the three ICF schemes used in this paper parameterized by O$^+$/O. The inspection of the three plots indicates that there are not substantial discrepancies or trends between the different ICF schemes.  The mean value of the differences is almost zero in all the cases,  implying the absence of systematic offsets. The comparison between the N abundances determined with the ICF(N$^+$) of \citet{peimbertcostero69} and \citet{mathisrosa91} does not show any clear trend. The comparison of \citet{izotovetal06} scheme with the other two indicates that this ICF(N$^+$) tend to give slightly lower N/H ratios in the higher ionised nebulae, those with lower values of the O$^+$/O ratio. In any case, this apparent trend, if real, is very subtle.

%%%%%%%%%%% 
%%%%%%%%%%% 

%%%%%%%%%%% 
\section{An empirical ionisation factor for N} 
\label{sec:icfN} 
%%%%%%%%%%%

\begin{figure} 
\centering 
\includegraphics[scale=0.37]{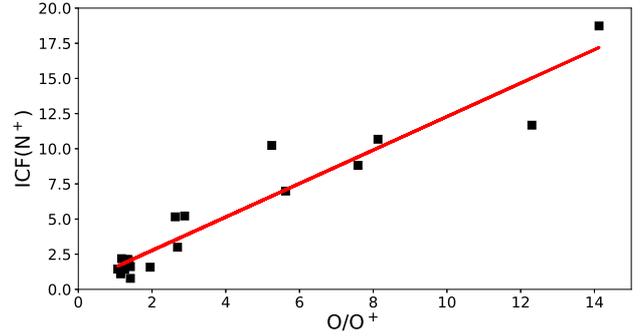} 
\caption{Values of our empirical ICF(N$^+$) as a function of the O/O$^{+}$ ratio for the {\hii} regions for which we need to use an ICF(N$^+$) to derive their N abundance -- those of the ICF group. The continuous red line represents a linear least-squares fit to the data. } 
\label{fig:ICFN} 
\end{figure}

We can formulate an empirical ICF(N$^+$) comparing the observed N$^+$/H$^+$ ratios of the objects of the ICF group and the total N abundance obtained using Eq.~\ref{eq:1} at their corresponding $R_{\rm G}$. The way we calculate the ICF(N$^+$) is the following: firstly we estimate the N/H ratio of each object that corresponds to its $R_{\rm G}$ using the radial gradient of N given by Eq.~\ref{eq:1}. Secondly, we divide the N/H ratio by the N$^+$/H$^+$ ratio observed for the object.  Finally, we calculate the linear least-squares fit of that ratio with respect to O/O$^{+}$. In Eq.~\ref{eq:6} and Figure~\ref{fig:ICFN} we show numerically and graphically, respectively, the parameters of the fit.

\begin{equation}\label{eq:6} \mathrm{ICF}(\mathrm{N}^{+}) = 0.39 + 1.19 \times \mathrm{O}/\mathrm{O}^{+}. \end{equation}

We estimate an uncertainty of about 0.10 dex for this empirical ICF(N$^+$). The range of validity of this relation correspond to the ionisation range covered by our objects: 1 $<$ O/O$^+$ $<$ 14. It must be taken into account that this empirical ICF(N$^+$) is based on real data of {\hii} regions, assumes our radial N gradient (given by Eq.~\ref{eq:1}) and that gaseous N abundances are well mixed in the Galaxy. As it is discussed in Sect~\ref{sec:Ngradient}, the use of this ICF(N$^+$) gives N abundances about 0.1 dex larger than the other schemes.

%%%%%%%%%%%%%%% 
%%%%%%%%%%%%%%%

%%%%%%%%%%% 
\section{The radial gradient of N/O} 
\label{sec:NOgradient} 
%%%%%%%%%%%

\begin{figure*} 
\centering 
\includegraphics[scale=0.78]{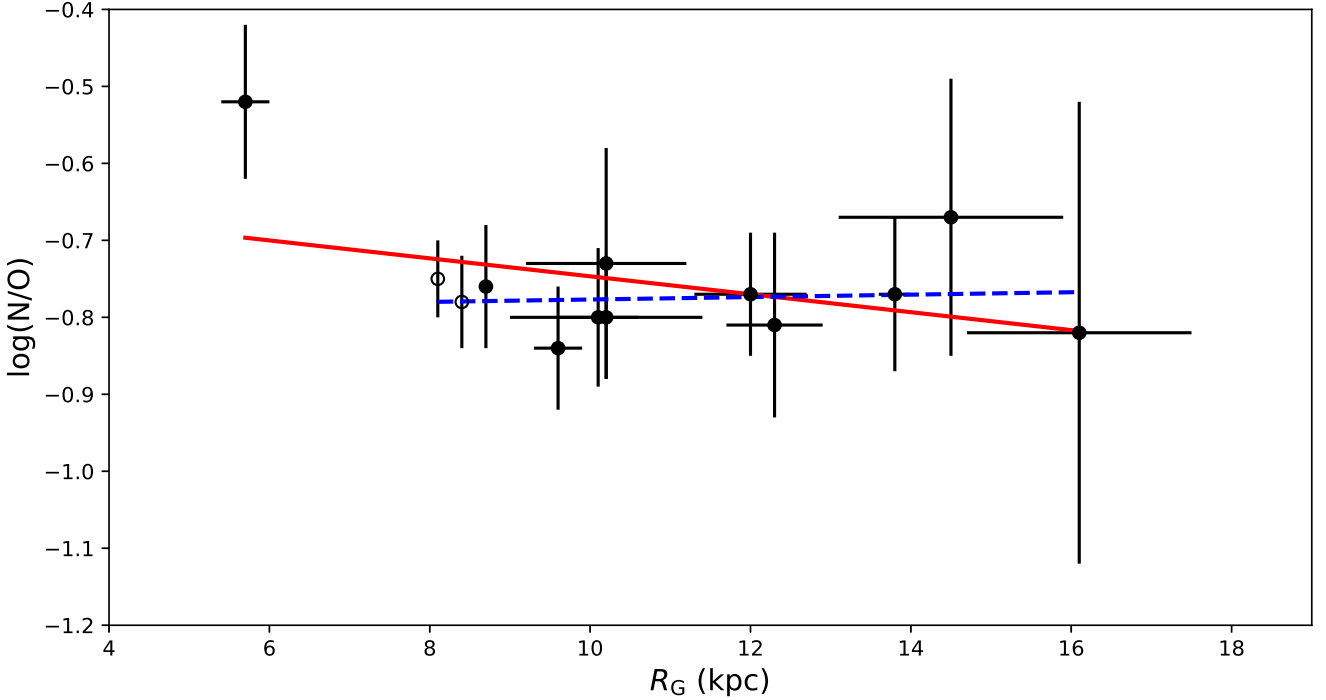} 
 \caption{Radial distribution of log(N/O) as a function of the Galactocentric distance, $R_{\rm G}$, for the Galactic {\hii} regions of the non-ICF group, for which we can assume N/H $\approx$ N$^+$/H$^+$. The empty circles correspond to IC~5146 and M~43, which data have been taken from the literature -- see text for references. The solid red line represents the least-squares fit to all the objects. The dashed blue line corresponds to the least-squares fit to the objects except the innermost one, Sh~2-61, that shows a much higher N/O ratio.} 
 \label{fig:NOgrad} 
 \end{figure*}

In Fig.~\ref{fig:NOgrad} we show the radial distribution of log(N/O) as a function of $R_{\rm G}$ for the Galactic {\hii} regions of the non-ICF group, for which we can assume N/H $\approx$ N$^+$/H$^+$. The least-squares linear fit to the data (represented by the red solid line in Fig.~\ref{fig:NOgrad}) is:

\begin{equation} \label{eq:7} \log(\mathrm{N/O}) = -0.63(\pm 0.17) - 0.012(\pm 0.018) R_\mathrm{G}, \end{equation}

\noindent which slope is rather small and can even be considered flat within the uncertainties. However, the value of the slope of the N/O gradient is very much dependent on the point corresponding to Sh~2-61, which is the innermost object of the non-ICF group and shows the highest N/O ratio. If we recalculate the least-squares linear fit excluding that object the slope becomes practically flat (represented by the blue dashed line in Fig.~\ref{fig:NOgrad}).

\begin{equation} \label{eq:8} \log(\mathrm{N/O}) = -0.79(\pm 0.22) + 0.002(\pm 0.021) R_\mathrm{G}. \end{equation}

\begin{figure} 
\centering 
\includegraphics[scale=0.37]{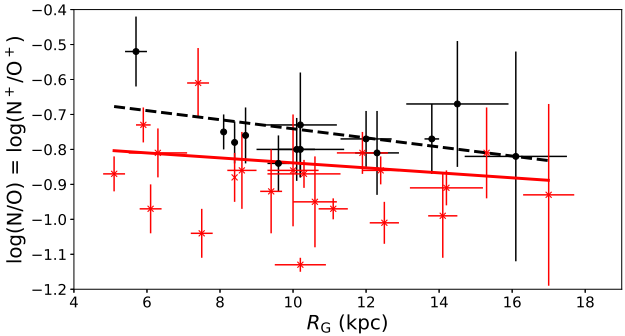} 
\caption{Radial distribution of log(N/O) as a function of the Galactocentric distance, $R_{\rm G}$, for all the Galactic {\hii} regions studied in this paper, which include objects of the non-ICF -- black circles -- and ICF groups -- red crosses. We have assumed that N/O = N$^+$/O$^+$, that implies the use of the ICF(N$^+$) of \citet{peimbertcostero69} for determining the N abundance. The solid red line represents the least-squares fit to all the objects. The dashed black line corresponds to the least-squares fit to the objects of the non-ICF group -- black circles -- given by Eq.~\ref{eq:7}.} 
\label{fig:NOgradall} 
\end{figure}

The flatness of the radial distribution of the N/O ratio of the {\hii} regions between 8 and 16 kpc, permits to assume a constant mean value of log(N/O) = $-$0.77 $\pm$ 0.04, at least in that range of $R_{\rm G}$. As we performed with the radial gradient of N in Section~\ref{sec:Ngradient}, we have also calculated the N/O gradient including the objects of the ICF group. In Fig.~\ref{fig:NOgradall} we show the N/O ratios of those nebulae using the ICF(N$^+$) of \citet{peimbertcostero69}, that implies to assume N/O = N$^+$/O$^+$. In this case, the least-squares linear fit to $R_{\rm G}$ and the N/O ratios gives:

\begin{equation} \label{eq:9} \log(\mathrm{N/O}) = -0.77(\pm 0.08) - 0.007(\pm 0.009) R_\mathrm{G}. \end{equation}

As in the previous fits shown in eqs.~\ref{eq:7} and \ref{eq:8}, the slope is again rather flat. We have also calculated the radial gradient of the N/O ratio using the ICF(N$^+$) of \citet{izotovetal06} or \citet{mathisrosa91} to determine the N abundance, but their slopes are the same: $-$0.009 $\pm$ 0.010 dex kpc$^{\rm -1}$ and in good agreement with that given by eq.~\ref{eq:9} assuming the ICF(N$^+$) of \citet{peimbertcostero69}. The mean value of log(N/O) we obtain using any of the three ICF(N$^+$) schemes is $-$0.84 $\pm$ 0.12, about 0.07 dex lower than the value we obtain for the objects of the non-ICF group.

%%%%%%%%%%%%%%% 
 %%%%%%%%%%%%%%%

%%%%%%%%%%%%%%% 
%%%%%%%%%%%%%%%

There are several determinations of the radial gradient of the N/O ratio based on {\hii} region spectra available in the literature. All of them have been obtained assuming an ICF scheme to determine the N abundance. Using optical data and the ICF(N$^+$) of \citet{peimbertcostero69}, \citet{shaveretal83} obtained a rather flat slope of $-$0.006 dex kpc$^{\rm -1}$, in good agreement with our results. Several determinations of N/O ratios in {\hii} regions use FIR spectra. Because {\fniii} lines are the only N lines observable in the FIR range of the spectra of ionised nebulae, an ICF(N$^{2+}$) is needed to derive the total N abundance in this case. With these kinds of observations, \citet{simpsonetal95} obtained a linear gradient of log(N/O) with a slope of $-$0.04 $\pm$ 0.01 dex kpc$^{\rm -1}$. However, they find a significantly better fit using a step fit with two levels, an inner a constant value of log(N/O) = $-$0.49 for objects located at $R_{\rm G}$ $<$ 6 kpc and and external one of log(N/O) = $-$0.74 for 6 kpc $<$ $R_{\rm G}$ $<$ 11 kpc. This last value of log(N/O) is very similar to the mean ratio we obtain from our optical data. \citet{simpsonetal95} used constant-density {\hii} region models to estimate the ICF(N$^{2+}$). \citet{rudolphetal06}, also based on FIR spectra and an ICF(N$^{2+}$) from photoionisation models, obtained a linear gradient of log(N/O) with a slope of $-$0.034 $\pm$ 0.006 dex kpc$^{\rm -1}$ for a sample of {\hii} regions located at $R_{\rm G}$ between 0 and 15 kpc. Inspecting the figure 7 of  \citet{rudolphetal06}, one can note that a two step fit similar to the one explored by \citet{simpsonetal95} seems also to reproduce the spatial distribution of the N/O ratios. Objects located at $R_{\rm G}$ > 6-7 kpc seem to show a fairly similar log(N/O) $\sim$ $-$0.74, while objects located in the inner zones show higher N/O ratios. The results based on FIR observations indicate an enhancement of the N/O ratio in the inner part of the Galactic disc but no strong evidence for an overall linear radial gradient of this quantity. As \citet{simpsonetal95} discuss in their paper, the uncertainties of the N/O ratios are larger for {\hii} regions with lower $T_{\rm eff}$ of their ionising stars. These nebulae have more uncertain ICF(N$^{2+}$) because their fractions of O$^+$ and N$^+$ are larger. Considering the positive gradient of {\elect} in the Galactic disc \citep[e.g.][]{afflerbachetal96, estebanetal17}, one would expect that N abundances based on FIR observations would be more affected by systematic errors in ICF(N$^{2+}$) in the case of {\hii} regions located in the inner parts of the Galaxy.

\citet{carigietal05}, from optical spectra of a small sample of {\hii} regions and assuming the ICF(N$^+$) of \citet{mathisrosa91}, calculated a radial gradient of log(N/O) of $-$0.042 $\pm$ 0.015 dex kpc$^{\rm -1}$. It is important to say that the spectra of the 8 {\hii} regions which data are used in \citet{carigietal05} are also included and reanalysed in our ICF group of objects. However, the N determinations used by \citet{carigietal05} are not comparable with ours because they use different sets of atomic data and assume {\ts} $>$ 0, which provide N abundances between 0.05 and 0.28 dex higher than our values. Moreover the range of $R_{\rm G}$ covered by their sample objects is between 5.1 and 11.1 kpc, quite narrower than our baseline.

Negative linear abundance gradients for N/O have been determined in many other spiral galaxies based on optical spectra. The vast majority of studies use an ICF(N$^+$) scheme and strong-line methods for estimating abundances because they lack direct determinations of {\elect}. Using published spectra of HII regions of 54 spiral galaxies and applying the $P$-method for estimating the O abundance, \citet{pilyuginetal04} have found negative slopes of the N/O gradient for all the galaxies. These authors use the approximation N/O $\approx$ N$^+$/O$^+$, therefore implicitly assuming the ICF(N$^+$) of \citet{peimbertcostero69}. In a more recent paper, \citet{belfioreetal17}, use resolved spectroscopic data from the Sloan Digital Sky Survey IV Mapping Nearby Galaxies at Apache Point Observatory (SDSS IV MaNGA) survey for 550 nearby galaxies, finding also negative log(N/O) gradients that seem to steepen with the stellar mass of the galaxies.  \citet{belfioreetal17} use the indirect N2O2 calibrator -- the formula given by \citet{pageletal92} -- for estimating the N/O ratio. This calibrator implicitly assumes  the ICF(N$^+$) of \citet{peimbertcostero69}. 
\citet{perezmonteroetal16} have analysed optical spectra of {\hii} regions in a sample of 350 spiral galaxies of the CALIFA survey, calculating O abundances using the semi-empirical routine H{\sc ii}-C{\sc hi-mistry}. Although they find that most galaxies show negative N/O gradients, a small fraction of them (about 4-10\%) display a flat or even a positive gradient. In this context, our result for the Milky Way seems to be not so rare.

The study of the evolution of N in galaxies is a difficult task. It can be produced in both massive and intermediate-mass stars, which enrich the interstellar gas on different timescales. The case of O is simpler because it is produced mostly by massive, short-lived stars. There is another complication with the nucleosynthesis of N. It needs the existence of previous C and O to be formed.  If N is produced from the C and O formed in the star itself then N is called primary but if the C and O were already existing at the time the star formed then it is called secondary. When the bulk of N is primary, its abundance increases in proportion to that of the O abundance and the N/O ratio remains constant as the O/H ratio increases \citep{talbotarnett74}. This behavior implies that the N yield is metallicity-independent. On the other hand, when N is secondary, its yield is metallicity-dependent and the N/O is no longer constant, increasing as the O/H increases \citep{clayton83}. The almost flat N/O gradient we find in this paper indicates that the bulk of the N is not formed by standard secondary processes at least in most part of the Galactic disc. \citet{israelianetal04} discuss the mechanisms that can produce primary N, which are the hot bottom burning (HBB) and rotational mixing (in intermediate mass stars) or rotation diffusion (in massive stars). Some authors have highlighted the difficulties to reconcile the observed and predicted N abundances in Galactic objects. For example,  the chemical evolution models of the Galactic disc of \citet{carigietal05} are successful in reproducing the O and C abundances but failed to fit the N/H values for metal-rich stars. In a recent paper, \citet{henryetal18} compare measured and model-predicted abundances of different elements including N in Galactic planetary nebulae (PN), finding a general discrepancy between the observed and model-predicted N/O ratios. These authors propose that an earlier-than-expected onset of HBB in AGB stars or the presence of extra mixing should be included in the stellar evolution models of low and intermediate mass stars to reconcile the predicted yields of N with observations. Considering all the discussion above, the inclusion of these aspects in evolution models  would probably promote the primary production of N in intermediate mass stars, which are responsible of a substantial fraction of the N enrichment in the Milky way.

%%%%%%%%%%% 
\section{A reassessment of the radial gradient of O} 
\label{sec:Ogradient} 
%%%%%%%%%%%

%%%%%%%%%%%%%%% 
\begin{figure*} 
\centering 
\includegraphics[scale=0.82]{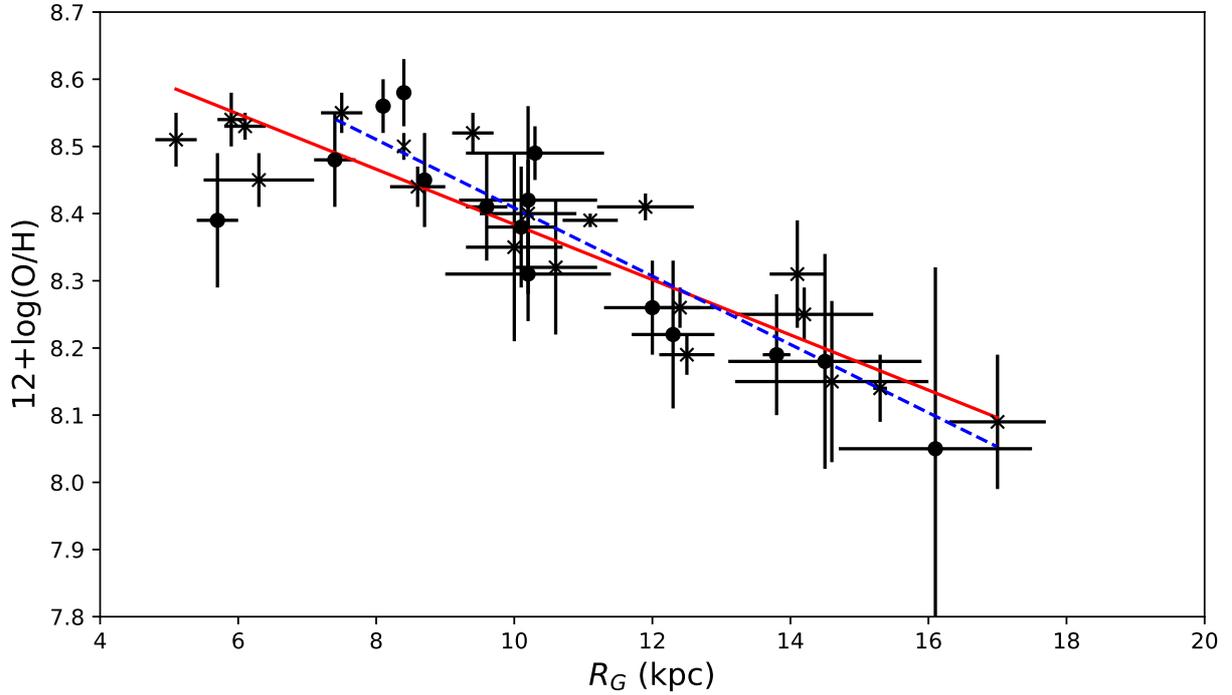} 
\caption{Radial distribution of the O abundance -- in units of 12+log(O/H) -- as a function of the Galactocentric distance, $R_{\rm G}$, for the Galactic {\hii} regions studied in this paper --  filled circles -- and those included in \citet{estebanetal17} -- crosses. The solid red line represents the least-squares fit to all objects. The dashed blue line corresponds to the least-squares fit to the {\hii} regions located at  $R_{\rm G}$ $>$ 7 kpc.} 
\label{fig:Ograd} 
\end{figure*} 
%%%%%%%%%%%%%%%

\citet{estebanetal17} presented O abundances of a sample of Galactic {\hii} regions based on direct determination of {\elect} and using the same methodology and atomic data than in the present paper. The combination of the results of the objects used in both works -- 35 objects -- represents the largest and best dataset available for estimating the radial O abundance gradient in the Milky Way. In fact, the spectra of all the objects except four have been taken with 10 m GTC or 8 m VLT telescopes. The spatial distribution of the O abundances is shown in Fig.~\ref{fig:Ograd}. The least-squares linear fit to the $R_{\rm G}$ and the O/H ratios including all the objects, gives the following radial O abundance gradient:

\begin{equation} \label{eq:10} 12 + \log(\mathrm{O/H}) = 8.80(\pm 0.09) - 0.041(\pm 0.006) R_\mathrm{G}; \end{equation}

\noindent for {\hii} regions in an interval of $R_{\rm G}$ extending from 5.1 to 17.0 kpc. This fit is also shown in Fig.~\ref{fig:Ograd} as a red continuous line. The gradient is almost identical to that obtained by \citet{estebanetal17}. It is worth noting that the O/H ratio of {\hii} regions located at $R_{\rm G}$ between 5.1 and about  8 kpc seem to break the general distribution of the rest of the objects. The distribution at this zone shows a flat or slightly positive gradient. This can be noted in figure 7 of \citet{estebanetal17}, but it was not reported because of the still small number of objects located at $R_{\rm G}$ $<$ 8 kpc studied in that paper. Our new results, specially the low O/H ratio of Sh~2-61, confirm this trend and make us to consider seriously the possibility of an inner drop or flattening of the Galactic radial O gradient at $R_{\rm G}$ $<$ 7-8 kpc. We have made a least-squares linear fit to the $R_{\rm G}$ and the O/H ratios but only including objects with $R_{\rm G}$ $>$ 7 kpc. In this case, the resulting radial O abundance gradient is:

\begin{equation} \label{eq:11} 12 + \log(\mathrm{O/H}) = 8.92(\pm 0.09) - 0.051(\pm 0.008) R_\mathrm{G}; \end{equation}

\noindent which is somewhat stepper than the fit we obtain for the whole sample. This second fit is indicated by a dashed blue line in Fig.~\ref{fig:Ograd}. 

The mean difference of the O abundance of the {\hii} regions represented in Fig.~\ref{fig:Ograd} and the abundance given by Eq.~\ref{eq:11} at their corresponding distance is $\pm$0.05 dex, of the order of the typical observational uncertainties. This is consistent with the results of \citet{estebanetal17} and indicates the homogeneity of the chemical composition of the Galactic ISM at a given radial distance. The maximum difference we find is $\pm$0.11 dex. This result contrast dramatically with the large scatter shown in figure 5 of \citet{rudolphetal06}. In fact, these last authors suggest that the spread in O or N 
abundances they found at a given $R_{\rm G}$ is real, indicating that the gas is not well mixed. Our results for both elements, O and N indicates just the contrary, the ISM seems to be fairly well mixed at a given distance along the Galactic disc. The high quality of our data, the homogeneous analysis and the proper cospatial determination of {\elect} for all the objects may be the reasons that explain 
these different conclusions. Recent studies on abundance gradients in other galaxies based on direct measurements of  {\elect} in {\hii} regions obtain slightly higher values of the scatter in O abundance, although they do not differ substantially from ours. \citet{croxalletal15, croxalletal16} obtain internal dispersions of 0.08 and 0.07 dex about the computed O gradient in NGC~5194 and M101 galaxies, respectively. However, other studies based on the application of strong-line methods \citep[e. g.][]{sanchezmenguianoetal16,hoetal17,vogtetal17} find dispersions of about 0.1-0.2 dex that have been attributed to azimuthal abundance variations along the galactic discs.

Let us come back to the apparent drop or flattening of the O/H ratio in the inner zones of the Galactic disc. This is a surprising result that may have important implications for Galactic chemical evolution models. Previous works on the abundance distribution of O, Fe and $\alpha$-elements in Cepheids in the inner part of the Galactic disc have revealed indications of a flat or even positive abundance gradient for $R_{\rm G}$ $<$ 5 kpc \citep{martinetal15, andrievskyetal16}. Moreover, metallicity gradients derived from SDSS-III/APOGEE observations of nearly 20,000 red giants by \citet{haydenetal14} indicate an apparent flattening at $R_{\rm G}$ $<$ 6 kpc, especially important for low-[$\alpha$/$Z$] stars. Although the flattening found with Cepheids or red giants seems to begin at somewhat smaller distances (at $R_{\rm G}$ $\sim$ 5-6 kpc) than suggested by {\hii} region observations, our results are qualitatively consistent, providing an independent confirmation of the presence of such change of slope of the gradients. \citet{andrievskyetal16} propose that this feature could be due to a decrease in the star formation rate produced by dynamic effects, possibly from the central Galactic bar. In a recent work, \citet{carigietal18} present chemical evolution models to reproduce the radial O gradient obtained by \citet{estebanetal17} and propose that the flattening of the O/H ratio in the inner zones may be due to an inside out quenching of the star formation history. Those authors find an abrupt decline of the star formation rate from 3 to 5 kpc, and from $\sim$ 4.1 to 8.4 Gyr. The origin of such quenching may be gas flows towards the Galactic Centre induced by the presence of the Galactic bar \citep{athanassoulaetal13, gavazzietal15}. \citet{haywoodetal16} and \citet{khoperskovetal18} propose another scenario of quenching without necessarily depleting the gas in the central parts. In this scenario, the action of a stellar bar increases the turbulence in the gas, preventing altogether the gas from collapsing and decreasing the star formation efficiency within the corotation radius of the bar. \citet{gerhard11} reviews and discusses the different determinations of the corotation radius of the Galactic bar, that go from 3.4 to 7 kpc. However, some dynamical models \citep{portailetal15, lietal16} constructed for reproducing recent density and kinematic data from red giants in the Galactic bulge/bar region \citep{wegggerhard13, kunderetal12} require a low pattern speed for the bar to reproduce the observations. This claim implies a corotation radius located at relatively large distances, $R_{\rm G}$ $\sim$ 6-7 kpc, a result that, although not coincident, is in qualitative agreement with our results for {\hii} regions that suggest the inner drop or flattening of the radial distribution of the O/H ratio would be located even more distant from the Galactic centre, at about 7-8 kpc.

The presence of an inner drop in the radial O abundance distribution of some spiral galaxies has been already found in several works \citep[e.g.][]{belleyroy92, rosalesortegaetal11, sanchezmenguianoetal17}. In all the cases, these features have been obtained from abundance analysis based on strong-line methods and not direct determinations of {\elect} of the {\hii} regions. In a very recent paper, \citet{sanchezmenguianoetal17} has presented MUSE@VLT data of about 100 spiral galaxies. These authors have found that about 35\% of the objects of their sample show an inner drop that appears at a very similar radial position for all the galaxies, about half of the effective radius, $R_{\rm e}$, of the galaxy. However, this prescription is not in agreement with the position of our inner drop for the Milky Way because it should be located at about $R_{\rm G}$ $\sim$ 2 kpc  \citep[considering that $R_{\rm e}$ is between 4-5 kpc for our Galaxy,][]{devaucouleurspence78, bovyrix13}. This value is much smaller than the approximate position of our change of slope at  $R_{\rm G}$ $\sim$ 7-8 kpc.

%%%%%%%%%%% 
\section{Conclusions} 
\label{sec:conclusions} 
%%%%%%%%%%%

We present very deep spectra of 13 Galactic {\hii} regions, with $R_{\rm G}$ between 5.7 and 16.1 kpc. The data have been obtained with the 10.4 m GTC telescope. We determine \elect({\fnii}) for all the objects using the direct method based on the detection of auroral lines. We have selected objects of very low ionisation degree for which the assumption N/H $\approx$ N$^+$/H$^+$ can be applied. This permits to calculate -- for the first time -- the radial abundance gradient of N avoiding the intrinsic uncertainty of assuming large ICFs to derive abundances. The slope of this new determination of the Galactic radial N abundance is $-$0.059 $\pm$ 0.009 dex kpc$^{-1}$. We have also calculated the N gradient including the sample of {\hii} regions with higher ionised degree studied by \citet{estebanetal17} (as well as two {\hii} regions analysed in this paper that do not show a very low ionisation degree: Sh~2-90 and Sh~2-152) and assuming three different ICF(N$^+$) schemes from the literature and the slope becomes somewhat less stepper, between $-$0.047 and $-$0.050 dex kpc$^{-1}$ but rather independent on the ICF scheme used. We find that N abundances determined assuming an ICF(N$^+$) scheme seem to be about 0.10 dex lower than those estimated from the radial gradient derived without the use of an ICF. We propose an empirical ICF(N$^+$) to estimate the total N abundance formulated from the comparison of the observed N$^+$/H$^+$ ratio of a given object and the N abundance estimated from our gradient at the corresponding $R_{\rm G}$ of the object.

We have studied the behavior of the radial distribution of the N/O finding that it is almost flat. A result that was also previously found by \citet{shaveretal83}. This indicates that the bulk of N is not formed by standard secondary processes. The mean N/O ratio along the Galactic disc is about $-$0.77 $\pm$ 0.04 dex.

We have made a reassessment of the radial O abundance gradient combining our results with the ones by \citet{estebanetal17}, who calculate the Galactic radial O abundance gradient based on deep optical spectra of 21 {\hii} regions analysed with the same methodology and atomic data that have been used in this paper. With this extended sample, we have homogeneous data of 35 {\hii} regions with direct determinations of {\elect} that we have used to study the Galactic radial O gradient from $R_{\rm G}$ between 5.1 and 17.0 kpc. We report the possible presence of a flattening or drop of the O abundance in the inner part of the Galactic disc, at $R_{\rm G}$ $<$ 7-8 kpc. This result confirms previous findings by \citet{martinetal15} and \citet{andrievskyetal16} from abundance data from Galactic Cepheids. This change of the slope of the radial O gradient may be produced by the quenching of star formation efficiency due to the dynamical action of the Galactic bar.

Finally, we find that the scatter of the N and O abundances of {\hii} regions with respect to our gradients is not substantially larger than the observational uncertainties -- typically about 0.05 dex -- indicating that both chemical elements are well mixed in the interstellar gas at a given distance along the Galactic disc.

\section*{Acknowledgements} We are very grateful to the constructive and detailed review carried out by our anonymous referee. His/her numerous and wise comments have helped to significantly improve this paper. We thank Manuel Peimbert, Jairo M\'endez-Abreu, Leticia Carigi and Rolf Kudritzki for their useful comments. This paper is based on observations made with the Gran Telescopio Canarias (GTC), installed in the Spanish Observatorio del Roque de los Muchachos of the Instituto de Astrof\'isica de Canarias, in the island of La Palma, Spain. This research has made use of Aladin Sky Atlas developed at CDS, Strasbourg, France. This work has been funded by the Spanish Ministerio de Econom\'ia y Competividad (MINECO) under project AYA2015-65205-P. JGR acknowledges support from an Advanced Fellowship from the Severo Ochoa excellence program (SEV-2015-0548).

%%%%%%%%%%%%%%%%%%%%%%%%%%%%%%%%%%%%%%%%%%%%%%%%%%



\bibliographystyle{mnras} 
\bibliography{cesar_bibliography} % if your bibtex file is called example.bib

%%%%%%%%%%%%%%%%%%%%%%%%%%%%%%%%%%%%%%%%%%%%%%%%%%

%%%%%%%%%%%%%%%%% APPENDICES %%%%%%%%%%%%%%%%%%%%%

\appendix

\section{Line intensity ratios}
\label{appex:1}
In this Appendix we include 3 tables with the line intensity ratios of the {\hii} regions of the sample. Each table contains the list of line identifications -- first 3 columns,  the reddening function, $f(\lambda)$ -- fourth column -- and dereddened line intensity ratios with respect to H$\beta$ for several objects -- rest of columns. Colons indicate line intensity errors of the order or greater than 40\%. The 2 last rows of each table include the reddening coefficient and the observed --uncorrected for reddening-- integrated H$\beta$ flux, $F$(H$\beta$), of the extracted aperture for each object. 

%Table of line intensities 1 
  \begin{table*}
  \centering
     \caption{Derredened line intensity ratios with respect to $I$(\hb) = 100 for Sh~2-61, Sh~2-152, Sh~2-235, Sh~2-271 and Sh~2-297.}
     \label{tab:lines_1}
    \begin{tabular}{lcccccccc}
     \hline                                              
        $\lambda_0$ & & & & \multicolumn{5}{c}{$I$($\lambda$)/$I$(\hb)} \\  
       (\AA) & Ion & ID& f($\lambda$) & Sh2-61 & Sh2-152 & Sh2-235 & Sh2-271& Sh2-297 \\ 
     \hline  
3726	&	{\foii}	&	1F	&	0.257	&	146.7	$\pm$	5.7	&	270.6	$\pm$	6.5	&	251	$\pm$	16	&	263.4	$\pm$	7.3	&	236.1	$\pm$	5.8	\\
3729	&	{\foii}	&	1F	&		&				&				&				&				&				\\
3750	&	{\hi}	&	H12	&	0.253	&	3.4	$\pm$	1.2	&	3.00	$\pm$	0.75	&	2.49	$\pm$	0.86	&	3.8	$\pm$	1.2	&	1.26	$\pm$	0.37	\\
3771	&	{\hi}	&	H11	&	0.249	&	2.89	$\pm$	0.73	&	4.09	$\pm$	0.41	&	3.7	$\pm$	1.3	&	4.1		:	&	1.63	$\pm$	0.63	\\
3798	&	{\hi}	&	H10	&	0.244	&	4.77	$\pm$	0.92	&	4.79	$\pm$	0.33	&	5.3	$\pm$	1.6	&	5.0	$\pm$	1.6	&	2.46	$\pm$	0.64	\\
3820	&	{\hei}	&	22	&	0.240	&		$-$		&	0.72		:	&		$-$		&				&		$-$		\\
3835	&	{\hi}	&	H9	&	0.237	&	4.5	$\pm$	1.1	&	6.85	$\pm$	0.31	&	7.7	$\pm$	1.2	&	6.4	$\pm$	1.6	&	4.0	$\pm$	1.1	\\
3867	&	{\hei}	&	20	&	0.231	&		$-$		&	0.57		:	&		$-$		&		$-$		&		$-$		\\
3869	&	{\fneiii} 	&	1F	&		&				&				&				&				&				\\
3889	&	{\hei}	&	5	&	0.227	&	8.8	$\pm$	1.2	&	15.57	$\pm$	0.46	&	14.8	$\pm$	1.1	&	11.9	$\pm$	1.2	&	9.7	$\pm$	1.3	\\
3889	&	{\hi}	&	H8	&		&				&				&				&				&				\\
3919	&	{\cii}	&	6	&	0.22	&		$-$		&	0.39	$\pm$	0.13	&		$-$		&		$-$		&		$-$		\\
3926	&	{\hei}	&	58	&	0.219	&		$-$		&	0.156	$\pm$	0.035	&		$-$		&		$-$		&		$-$		\\
3936	&	{\hei}	&	57	&	0.217	&		$-$		&		$-$		&		$-$		&	1.59	$\pm$	0.60	&		$-$		\\
3967	&	{\fneiii} 	&	1F	&	0.211	&	14.45	$\pm$	0.59	&	17.12	$\pm$	0.43	&	16.4	$\pm$	3.1	&	17.3	$\pm$	1.1	&	13.09	$\pm$	0.72	\\
3970	&	{\hi}	&	H7	&		&				&				&				&				&				\\
4009	&	{\hei}	&	55	&	0.202	&	2.2		:	&		$-$		&		$-$		&	1.32	$\pm$	0.39	&		$-$		\\
4026	&	{\hei}	&	18	&	0.198	&		$-$		&	1.57	$\pm$	0.14	&	1.0		:	&		$-$		&		$-$		\\
4069	&	{\fsii}	&	1F	&	0.189	&	5.4	$\pm$	1.1	&	1.58	$\pm$	0.15	&	1.92	$\pm$	0.37	&	1.50	$\pm$	0.41	&	1.23	$\pm$	0.49	\\
4076	&	{\fsii}	&	1F	&	0.187	&	1.6	  :	&	1.18	$\pm$	0.14	&	1.20	$\pm$	0.26	&		$-$		&		$-$		\\
4102	&	{\hi}	&	H6	&	0.182	&	24.4	$\pm$	1.2	&	25.93	$\pm$	0.55	&	26.29	$\pm$	0.81	&	25.53	$\pm$	0.80	&	23.04	$\pm$	0.69	\\
4267	&	{\cii}	&	6	&	0.144	&		$-$		&	0.104	$\pm$	0.038	&		$-$		&		$-$		&		$-$		\\
4340	&	{\hi}	&	H$\gamma$	&	0.127	&	45.4	$\pm$	1.2	&	47.21	$\pm$	0.96	&	46.1	$\pm$	1.0	&	45.2	$\pm$	1.0	&	44.1	$\pm$	1.0	\\
4363	&	{\foiii}	&	2F	&	0.121	&		$-$		&	0.113 :	&		$-$		&		$-$		&		$-$		\\
4388	&	{\hei}	&	51	&	0.115	&		$-$		&	0.360	$\pm$	0.080	&		$-$		&		$-$		&		$-$		\\
4471	&	{\hei}	&	14	&	0.096	&		$-$		&	3.512	$\pm$	0.088	&		$-$		&	0.85	$\pm$	0.15	&		$-$		\\
4639	&	{\oii}	&	1	&	0.055	&		$-$		&	0.051		:	&		$-$		&		$-$		&		$-$		\\
4658	&	{\ffeiii}	&	3F	&	0.050	&	0.41	$\pm$	0.15	&	0.712	$\pm$	0.029	&	0.297	$\pm$	0.085	&	0.79	$\pm$	0.15	&	0.372	$\pm$	0.083	\\
4702	&	{\ffeiii}	&	3F	&	0.039	&		$-$		&	0.207	$\pm$	0.032	&		$-$		&		$-$		&		$-$		\\
4713	&	{\hei}	&	12	&	0.036	&		$-$		&	0.426	$\pm$	0.051	&	0.166	$\pm$	0.054	&		$-$		&		$-$		\\
4734	&	{\ffeiii}	&	3F	&	0.031	&		$-$		&	0.061		:	&		$-$		&		$-$		&		$-$		\\
4755	&	{\ffeiii}	&	3F	&	0.026	&		$-$		&	0.144	$\pm$	0.034	&		$-$		&		$-$		&		$-$		\\
4770	&	{\ffeiii}	&	3F	&	0.023	&		$-$		&	0.080		:	&		$-$		&		$-$		&		$-$		\\
4815	&	{\ffeii}	&	20F	&	0.012	&	0.62	$\pm$	0.20	&		$-$		&		$-$		&		$-$		&		$-$		\\
4861	&	{\hi}	&	{\hb}	&	0.000	&	100.0	$\pm$	2.1	&	100.0	$\pm$	2.0	&	100.0	$\pm$	2.0	&	100.0	$\pm$	2.0	&	100.0	$\pm$	2.1	\\
4881	&	{\ffeiii}	&	3F	&	-0.005	&		$-$		&	0.183	$\pm$	0.048	&		$-$		&		$-$		&		$-$		\\
4922	&	{\hei}	&	48	&	-0.015	&		$-$		&	0.888	$\pm$	0.060	&	0.540	$\pm$	0.095	&	0.200	$\pm$	0.080	&		$-$		\\
4959	&	{\foiii}	&	1F	&	-0.024	&		$-$		&	14.41	$\pm$	0.30	&	1.846	$\pm$	0.083	&	0.40	$\pm$	0.10	&	0.90		:	\\
4986	&	{\ffeiii}	&	3F	&	-0.031	&		$-$		&	0.180	$\pm$	0.058	&	0.244	$\pm$	0.091	&	0.56	$\pm$	0.13	&	0.42	$\pm$	0.12	\\
5007	&	{\foiii}	&	1F	&	-0.036	&	0.27	:	&	43.66	$\pm$	0.98	&	5.73	$\pm$	0.26	&	0.83	$\pm$	0.15	&	2.70	$\pm$	0.23	\\
5016	&	{\hei}	&	4	&	-0.038	&		$-$		&	1.272	$\pm$	0.057	&	1.032	$\pm$	0.065	&	0.54	$\pm$	0.14	&		$-$		\\
5041	&	\ion{Si}{ii}	&	5	&	-0.044	&	0.353	$\pm$	0.090	&	0.0613	$\pm$	0.0025	&	0.354	$\pm$	0.090	&		$-$		&	0.194	$\pm$	0.072	\\
5048	&	{\hei}	&	47	&	-0.046	&		$-$		&	0.1126	$\pm$	0.0041	&		$-$		&		$-$		&		$-$		\\
5056	&	\ion{Si}{ii}	&	5	&	-0.048	&	0.96	$\pm$	0.17	&	0.1769	$\pm$	0.0064	&	0.230	$\pm$	0.039	&		$-$		&	0.224	$\pm$	0.051	\\
5159	&	{\ffeii}	&	19F	&	-0.073	&	0.43	$\pm$	0.16	&		$-$		&		$-$		&		$-$		&		$-$		\\
5198	&	{\fni}	&	1F	&	-0.082	&	5.00	$\pm$	0.27	&	0.317	$\pm$	0.028	&	0.510	$\pm$	0.087	&	0.809	$\pm$	0.079	&	1.68	$\pm$	0.16	\\
5262	&	{\ffeii}	&	19F	&	-0.098	&	0.288	$\pm$	0.087	&		$-$		&		$-$		&		$-$		&	0.11		:	\\
5270	&	{\ffeiii}	&	1F	&	-0.100	&	0.497	$\pm$	0.050	&	0.343	$\pm$	0.033	&	0.123	$\pm$	0.037	&	0.439	$\pm$	0.073	&	0.246	$\pm$	0.086	\\
5412	&	{\heii}	&	4.7	&	-0.134	&		$-$		&		$-$		&		$-$		&	0.093		:	&		$-$		\\
5518	&	{\fcliii}	&	1F	&	-0.154	&		$-$		&	0.429	$\pm$	0.017	&	0.285	$\pm$	0.051	&	0.301	$\pm$	0.063	&	0.222	$\pm$	0.052	\\
5538	&	{\fcliii}	&	1F	&	-0.158	&		$-$		&	0.363	$\pm$	0.021	&	0.159	$\pm$	0.029	&	0.212	$\pm$	0.060	&	0.128	$\pm$	0.045	\\
5755	&	{\fnii}	&	3F	&	-0.194	&	1.07	$\pm$	0.15	&	0.896	$\pm$	0.023	&	0.772	$\pm$	0.050	&	1.019	$\pm$	0.076	&	0.842	$\pm$	0.067	\\
5876	&	{\hei}	&	11	&	-0.215	&		$-$		&	9.25	$\pm$	0.19	&	5.73	$\pm$	0.15	&	2.36	$\pm$	0.16	&	2.34	$\pm$	0.30	\\
5958	&	\ion{Si}{ii}	&	4	&	-0.228	&	0.401	$\pm$	0.077	&	0.065	$\pm$	0.013	&		$-$		&		$-$		&	0.141	$\pm$	0.052	\\
5979	&	\ion{Si}{ii}	&	4	&	-0.231	&	0.612	$\pm$	0.092	&	0.104	$\pm$	0.016	&	0.115	$\pm$	0.035	&	0.099	$\pm$	0.036	&	0.171	$\pm$	0.041	\\
6046	&	{\oi}	&	22	&	-0.242	&	0.619	$\pm$	0.067	&	0.023		:	&		$-$		&		$-$		&		$-$		\\
6300	&	{\foi}	&	1F	&	-0.282	&	4.31	$\pm$	0.17	&	0.458	$\pm$	0.053	&	1.167	$\pm$	0.071	&	1.530	$\pm$	0.064	&	3.02	$\pm$	0.13	\\
6312	&	{\fsiii}	&	3F	&	-0.283	&		$-$		&	1.298	$\pm$	0.059	&	0.637	$\pm$	0.054	&	0.575	$\pm$	0.079	&		$-$		\\
6347	&	\ion{Si}{ii}	&	4	&	-0.289	&	0.74	$\pm$	0.17	&	0.152	$\pm$	0.058	&	0.129	$\pm$	0.046	&		$-$		&		$-$		\\
6364	&	{\foi}	&	1F	&	-0.291	&	1.25	$\pm$	0.14	&	0.14		:	&	0.379	$\pm$	0.075	&	0.502	$\pm$	0.072	&	0.82	$\pm$	0.12	\\
\end{tabular}
\end{table*}
%%%%%%%%%%%%%%%%%%%%%
\setcounter{table}{0}
   \begin{table*}
  \centering
     \caption{continued}
    \begin{tabular}{lccccccccc}
     \hline  
        $\lambda_0$ & & & & \multicolumn{5}{c}{$I$($\lambda$)/$I$(\hb)} \\  
       (\AA) & Ion & ID& f($\lambda$) & Sh2-61 & Sh2-152 & Sh2-235 & Sh2-271& Sh2-297 \\ 
     \hline  
6371	&	\ion{Si}{ii}	&	2	&	-0.292	&	0.292	$\pm$	0.080	&	0.080		:	&		$-$		&		$-$		&		$-$		\\
6548	&	{\fnii}	&	1F	&	-0.318	&	53.3	$\pm$	1.6	&	33.85	$\pm$	0.70	&	31.28	$\pm$	0.83	&	33.7	$\pm$	1.0	&	38.47	$\pm$	0.94	\\
6563	&	{\hi}	&	{\ha}	&	-0.320	&	279.8	$\pm$	8.5	&	293.4	$\pm$	6.1	&	268.2	$\pm$	6.6	&	297.4	$\pm$	7.4	&	292.4$^{\rm a}$	$\pm$	7.1	\\
6583	&	{\fnii}	&	1F	&	-0.323	&	163.0	$\pm$	5.0	&	104.0	$\pm$	2.2	&	95.5	$\pm$	2.4	&	100.8	$\pm$	2.5	&	117.6	$\pm$	2.9	\\
6678	&	{\hei}	&	46	&	-0.336	&		$-$		&	2.555	$\pm$	0.057	&	1.556	$\pm$	0.062	&	0.595	$\pm$	0.090	&	0.553	$\pm$	0.070	\\
6716	&	{\fsii}	&	2F	&	-0.342	&	37.0	$\pm$	1.2	&	12.96	$\pm$	0.27	&	25.99	$\pm$	0.66	&	24.41	$\pm$	0.57	&	32.23	$\pm$	0.80	\\
6731	&	{\fsii}	&	2F	&	-0.344	&	47.7	$\pm$	1.5	&	14.78	$\pm$	0.31	&	20.42	$\pm$	0.52	&	18.58	$\pm$	0.44	&	24.88	$\pm$	0.62	\\
7002	&	{\oi}	&	21	&	-0.379	&		$-$		&	0.043	$\pm$	0.013	&		$-$		&		$-$		&		$-$		\\
7065	&	{\hei}	&	10	&	-0.387	&		$-$		&	3.42	$\pm$	0.24	&	2.15	$\pm$	0.17	&	2.06	$\pm$	0.24	&	5.59	$\pm$	0.19	\\
7136	&	{\fariii}	&	1F	&	-0.396	&		$-$		&	7.84	$\pm$	0.55	&	2.98	$\pm$	0.22	&	0.76	$\pm$	0.12	&	1.26	$\pm$	0.11	\\
7155	&	{\ffeii}	&	14F	&	-0.399	&	0.30	$\pm$	0.07	&				&		$-$		&		$-$		&		$-$		\\
7231	&	{\cii}	&	3	&	-0.408	&		$-$		&	0.080	$\pm$	0.022	&		$-$		&		$-$		&		$-$		\\
7236	&	{\cii}	&	3	&	-0.409	&		$-$		&	0.139	$\pm$	0.036	&	0.40		:	&		$-$		&		$-$		\\
7281	&	{\hei}	&	45	&	-0.414	&		$-$		&	0.464	$\pm$	0.071	&		$-$		&		$-$		&		$-$		\\
7318	&	{\foii}	&	2F	&	-0.418	&	1.68	$\pm$	0.13	&	3.36	$\pm$	0.24	&	1.57	$\pm$	0.12	&	2.00	$\pm$	0.23	&	1.53	$\pm$	0.11	\\
7320	&	{\foii}	&	2F	&		&				&				&				&				&				\\
7330	&	{\foii}	&	2F	&	-0.420	&	1.31	$\pm$	0.10	&	3.01	$\pm$	0.21	&	1.329	$\pm$	0.097	&	1.52	$\pm$	0.17	&	1.294	$\pm$	0.096	\\
7331	&	{\foii}	&	2F	&		&				&				&				&				&				\\
7751	&	{\fariii}	&	2F	&	-0.467	&		$-$		&	1.65	$\pm$	0.13	&	1.19	$\pm$	0.16	&		$-$		&		$-$		\\
\multicolumn{4}{l}{$c$(H$\beta$)} & 1.78 $\pm$ 0.03 & 1.63 $\pm$ 0.01 & 1.39 $\pm$ 0.02 &  1.16 $\pm$ 0.01 &  0.25 $\pm$ 0.01 \\
\multicolumn{4}{l}{$F$(H$\beta$)$^{\rm b}$} & 2.01 $\pm$ 0.04 & 17.1 $\pm$ 0.3 & 5.1 $\pm$ 0.1 & 3.20 $\pm$ 0.06 & 19.5 $\pm$ 0.4 \\
     \hline                                              
    \end{tabular}
    \begin{description}
      \item[$^{\rm a}$] Line flux from short exposure spectrum. Line saturated in long exposure spectrum.   
      \item[$^{\rm b}$] Flux uncorrected for reddening in units of 10$^{-14}$ erg cm$^{-2}$ s$^{-1}$.                   
    \end{description}
   \end{table*}		
%%%%%%%%%%%%%%%%

%Table of line intensities 2 
  \begin{table*}
  \centering
     \caption{Derredened line intensity ratios with respect to $I$(\hb) = 100 for Sh~2-90, Sh~2-219, Sh~2-237, Sh~2-257 and Sh~2-266.}
     \label{tab:lines_2}
    \begin{tabular}{lcccccccc}
     \hline                                              
        $\lambda_0$ & & & & \multicolumn{5}{c}{$I$($\lambda$)/$I$(\hb)} \\  
       (\AA) & Ion & ID& f($\lambda$) & Sh2-90 & Sh2-219 & Sh2-237 & Sh2-257& Sh2-266 \\ 
     \hline  
3726	&	{\foii}	&	1F	&	0.257	&	233	$\pm$	15	&	259	$\pm$	14	&	295.3	$\pm$	8.6	&	220.4	$\pm$	8.1	&	166.7	$\pm$	8.3	\\
3729	&	{\foii}	&	1F	&		&				&				&				&				&				\\
3750	&	{\hi}	&	H12	&	0.253	&		$-$		&		$-$		&		$-$		&	2.4		:	&		$-$		\\
3771	&	{\hi}	&	H11	&	0.249	&	8.8		:	&		$-$		&		$-$		&	1.9		:	&		$-$		\\
3798	&	{\hi}	&	H10	&	0.244	&		$-$		&		$-$		&		$-$		&	4.5	$\pm$	1.4	&		$-$		\\
3835	&	{\hi}	&	H9	&	0.237	&		$-$		&	10		:	&		$-$		&	7.1	$\pm$	1.4	&	11.6	$\pm$	4.4	\\
3889	&	{\hei}	&	5	&	0.227	&	17		:	&	11.8	$\pm$	2.7	&		$-$		&	10.8	$\pm$	1.2	&	12		:	\\
3889	&	{\hi}	&	H8	&		&				&				&				&				&				\\
3967	&	{\fneiii} 	&	1F	&	0.211	&	22.8	$\pm$	3.8	&	14.4	$\pm$	1.8	&		$-$		&	13.6	$\pm$	1.6	&	17.8	$\pm$	4.7	\\
3970	&	{\hi}	&	H7	&		&				&				&				&				&				\\
4069	&	{\fsii}	&	1F	&	0.189	&		$-$		&		$-$		&		$-$		&	2.0		:	&		$-$		\\
4102	&	{\hi}	&	H6	&	0.182	&	26.5	$\pm$	3.4	&	26.0	$\pm$	1.6	&	18.9	$\pm$	1.0	&	23.8	$\pm$	1.0	&	28.4	$\pm$	4.9	\\
4340	&	{\hi}	&	H$\gamma$	&	0.127	&	47.0	$\pm$	2.2	&	45.4	$\pm$	2.0	&	44.1	$\pm$	1.6	&	45.6	$\pm$	1.3	&	41.6	$\pm$	2.3	\\
4471	&	{\hei}	&	14	&	0.096	&	4.41	$\pm$	0.75	&		$-$		&		$-$		&	0.41		:	&	1.4		:	\\
4658	&	{\ffeiii}	&	3F	&	0.050	&		$-$		&		$-$		&		$-$		&		$-$		&	1.54	$\pm$	0.45	\\
4861	&	{\hi}	&	{\hb}	&	0.000	&	100.0	$\pm$	2.3	&	100.0	$\pm$	2.3	&	100.0	$\pm$	2.2	&	100.0	$\pm$	2.0	&	100.0	$\pm$	2.2	\\
4922	&	{\hei}	&	48	&	-0.015	&	1.40	$\pm$	0.54	&		$-$		&		$-$		&		$-$		&		$-$		\\
4959	&	{\foiii}	&	1F	&	-0.024	&		$-$		&		$-$		&	0.91	$\pm$	0.30	&	0.65	$\pm$	0.19	&		$-$		\\
4986	&	{\ffeiii}	&	3F	&	-0.031	&		$-$		&		$-$		&		$-$		&		$-$		&	0.82	$\pm$	0.33	\\
5007	&	{\foiii}	&	1F	&	-0.036	&	75.0	$\pm$	1.7	&	1.2		:	&	2.92	$\pm$	0.36	&	2.24	$\pm$	0.10	&	1.01	$\pm$	0.34	\\
5016	&	{\hei}	&	4	&	-0.038	&	1.55	$\pm$	0.27	&		$-$		&		$-$		&	0.372	$\pm$	0.075	&		$-$		\\
5041	&	\ion{Si}{ii}	&	5	&	-0.044	&		$-$		&		$-$		&		$-$		&	0.332	$\pm$	0.074	&		$-$		\\
5056	&	\ion{Si}{ii}	&	5	&	-0.048	&		$-$		&		$-$		&		$-$		&	0.28		:	&		$-$		\\
5159	&	{\ffeii}	&	19F	&	-0.073	&		$-$		&		$-$		&		$-$		&		$-$		&	0.62		:	\\
5198	&	{\fni}	&	1F	&	-0.082	&	0.78	$\pm$	0.17	&	1.17	$\pm$	0.32	&	3.85	$\pm$	0.32	&	1.28	$\pm$	0.14	&	6.63	$\pm$	0.94	\\
5262	&	{\ffeii}	&	19F	&	-0.098	&		$-$		&		$-$		&	0.23		:	&		$-$		&		$-$		\\
5270	&	{\ffeiii}	&	1F	&	-0.100	&		$-$		&		$-$		&	0.344	$\pm$	0.098	&	0.105	$\pm$	0.026	&	0.77	$\pm$	0.23	\\
5518	&	{\fcliii}	&	1F	&	-0.154	&	0.47	$\pm$	0.10	&	0.37		:	&	0.41	$\pm$	0.15	&	0.207	$\pm$	0.048	&		$-$		\\
5538	&	{\fcliii}	&	1F	&	-0.158	&	0.59	$\pm$	0.10	&		$-$		&		$-$		&	0.109	$\pm$	0.032	&		$-$		\\
5755	&	{\fnii}	&	3F	&	-0.194	&	1.27	$\pm$	0.10	&	0.870	$\pm$	0.087	&	1.152	$\pm$	0.092	&	0.732	$\pm$	0.069	&	0.83	$\pm$	0.12	\\
5876	&	{\hei}	&	11	&	-0.215	&	12.05	$\pm$	0.38	&	3.30	$\pm$	0.29	&	0.66	$\pm$	0.11	&	2.73	$\pm$	0.18	&	1.32	$\pm$	0.23	\\
6046	&	{\oi}	&	22	&	-0.242	&		$-$		&		$-$		&	0.281	$\pm$	0.088	&		$-$		&		$-$		\\
6300	&	{\foi}	&	1F	&	-0.282	&	2.47	$\pm$	0.40	&	0.87	$\pm$	0.21	&	2.08	$\pm$	0.19	&		$-$		&	12.5	$\pm$	1.3	\\
6312	&	{\fsiii}	&	3F	&	-0.283	&		$-$		&		$-$		&		$-$		&	0.64	$\pm$	0.24	&		$-$		\\
6364	&	{\foi}	&	1F	&	-0.291	&	0.60	$\pm$	0.15	&		$-$		&	0.60	$\pm$	0.15	&		$-$		&	3.45	$\pm$	0.67	\\
6548	&	{\fnii}	&	1F	&	-0.318	&	49.3	$\pm$	1.2	&	28.73	$\pm$	0.78	&	36.06	$\pm$	0.88	&	31.5	$\pm$	1.2	&	31.0	$\pm$	2.8	\\
6563	&	{\hi}	&	{\ha}	&	-0.320	&	297.4	$\pm$	7.3	&	288.5	$\pm$	7.8	&	289.0$^{\rm a}$	$\pm$	6.4	&	278	$\pm$	10	&	294	$\pm$	26	\\
6583	&	{\fnii}	&	1F	&	-0.323	&	151.3	$\pm$	3.8	&	88.9	$\pm$	2.4	&	114.4	$\pm$	2.5	&	96.3	$\pm$	3.6	&	93.2	$\pm$	8.5	\\
6678	&	{\hei}	&	46	&	-0.336	&	3.28	$\pm$	0.14	&	0.83	$\pm$	0.15	&		$-$		&	0.683	$\pm$	0.050	&	0.34	$\pm$	0.13	\\
6716	&	{\fsii}	&	2F	&	-0.342	&	26.50	$\pm$	0.72	&	31.53	$\pm$	0.85	&	34.59	$\pm$	0.92	&	37.9	$\pm$	1.5	&	46.9	$\pm$	4.5	\\
6731	&	{\fsii}	&	2F	&	-0.344	&	21.39	$\pm$	0.58	&	22.60	$\pm$	0.64	&	32.96	$\pm$	0.75	&	29.3	$\pm$	1.2	&	42.8	$\pm$	4.1	\\
7065	&	{\hei}	&	10	&	-0.387	&	2.16	$\pm$	0.23	&	2.53	$\pm$	0.32	&	1.88	$\pm$	0.23	&	1.90	$\pm$	0.16	&		$-$		\\
7136	&	{\fariii}	&	1F	&	-0.396	&	9.02	$\pm$	0.66	&	1.12	$\pm$	0.16	&	0.87	$\pm$	0.12	&	1.15	$\pm$	0.11	&		$-$		\\
7281	&	{\hei}	&	45	&	-0.414	&		$-$		&		$-$		&				&		$-$		&		$-$		\\
7318	&	{\foii}	&	2F	&	-0.418	&	1.89	$\pm$	0.22	&	1.86	$\pm$	0.32	&	2.27	$\pm$	0.23	&	1.21	$\pm$	0.12	&	3.64	$\pm$	0.54	\\
7320	&	{\foii}	&	2F	&		&				&				&				&				&				\\
7330	&	{\foii}	&	2F	&	-0.420	&	1.72	$\pm$	0.20	&	1.66	$\pm$	0.43	&	2.20	$\pm$	0.22	&	0.921	$\pm$	0.088	&	3.15	$\pm$	0.43	\\
7331	&	{\foii}	&	2F	&		&				&				&				&				&				\\
7751	&	{\fariii}	&	2F	&	-0.467	&	2.96	$\pm$	0.53	&		$-$		&		$-$		&		$-$		&		$-$		\\
\multicolumn{4}{l}{$c$(H$\beta$)} & 2.19 $\pm$ 0.02 & 1.00 $\pm$ 0.02 & 1.24 $\pm$ 0.01 &  1.06 $\pm$ 0.04 &  1.53 $\pm$ 0.12 \\
\multicolumn{4}{l}{$F$(H$\beta$)$^{\rm b}$} & 1.17 $\pm$ 0.02 & 1.08 $\pm$ 0.02 & 2.03 $\pm$ 0.02 & 2.91 $\pm$ 0.06 & 0.31 $\pm$ 0.4 \\
     \hline                                              
    \end{tabular}
    \begin{description}
      \item[$^{\rm a}$] Line flux from short exposure spectrum. Line saturated in long exposure spectrum.   
      \item[$^{\rm b}$] Flux uncorrected for reddening in units of 10$^{-14}$ erg cm$^{-2}$ s$^{-1}$.                   
    \end{description}
   \end{table*}	
%%%%%%%%%%%%%%%%

%Table of line intensities 3 
  \begin{table*}
  \centering
     \caption{Derredened line intensity ratios with respect to $I$(\hb) = 100 for Sh~2-175, Sh~2-270 and Sh~2-285.}
     \label{tab:lines_3}
    \begin{tabular}{lcccccc}
     \hline                                              
        $\lambda_0$ & & & & \multicolumn{3}{c}{$I$($\lambda$)/$I$(\hb)} \\  
       (\AA) & Ion & ID& f($\lambda$) & Sh2-175 & Sh2-270& Sh2-285 \\ 
     \hline
3726	&	{\foii}	&	1F	&	0.257	&	144	$\pm$	10	&	226	$\pm$	26	&	213.3	$\pm$	8.0	\\
3729	&	{\foii}	&	1F	&		&				&				&				\\
3771	&	{\hi}	&	H11	&	0.249	&		$-$		&		$-$		&	3.48	$\pm$	0.99	\\
3798	&	{\hi}	&	H10	&	0.244	&		$-$		&		$-$		&	4.8	$\pm$	1.6	\\
3835	&	{\hi}	&	H9	&	0.237	&		$-$		&		$-$		&	5.6	$\pm$	1.2	\\
3889	&	{\hei}	&	5	&	0.227	&	17.9	$\pm$	4.3	&		$-$		&	9.4	$\pm$	1.0	\\
3889	&	{\hi}	&	H8	&		&				&				&				\\
3967	&	{\fneiii} 	&	1F	&	0.211	&	22.9	$\pm$	4.6	&		$-$		&	15.25	$\pm$	0.97	\\
3970	&	{\hi}	&	H7	&		&				&				&				\\
4069	&	{\fsii}	&	1F	&	0.189	&		$-$		&		$-$		&	2.34	$\pm$	0.62	\\
4076	&	{\fsii}	&	1F	&	0.187	&		$-$		&		$-$		&	0.38	$\pm$	0.13	\\
4102	&	{\hi}	&	H6	&	0.182	&	23.07	$\pm$	2.57	&	24.1	$\pm$	1.8	&	24.4	$\pm$	1.0	\\
4340	&	{\hi}	&	H$\gamma$	&	0.127	&	42.1	$\pm$	2.4	&	43.3	$\pm$	2.7	&	45.3	$\pm$	1.2	\\
4861	&	{\hi}	&	{\hb}	&	0.000	&	100.0	$\pm$	2.2	&	100.0	$\pm$	2.5	&	100.0	$\pm$	2.0	\\
5041	&	\ion{Si}{ii}	&	5	&	-0.044	&	1.5		:	&	1.90	$\pm$	0.46	&		$-$		\\
5056	&	\ion{Si}{ii}	&	5	&	-0.048	&	0.84		:	&	3.11	$\pm$	0.65	&		$-$		\\
5198	&	{\fni}	&	1F	&	-0.082	&	2.69	$\pm$	0.45	&	8.37	$\pm$	0.56	&	2.33	$\pm$	0.15	\\
5262	&	{\ffeii}	&	19F	&	-0.098	&		$-$		&		$-$		&	0.13	$\pm$	0.04	\\
5755	&	{\fnii}	&	3F	&	-0.194	&	0.452	$\pm$	0.054	&	0.90	$\pm$	0.22	&	0.820	$\pm$	0.082	\\
6300	&	{\foi}	&	1F	&	-0.282	&	1.81	$\pm$	0.22	&	1.88	$\pm$	0.40	&		$-$		\\
6347	&	\ion{Si}{ii}	&	4	&	-0.289	&	0.57	$\pm$	0.19	&	1.3		:	&		$-$		\\
6364	&	{\foi}	&	1F	&	-0.291	&	0.43	$\pm$	0.11	&		$-$		&		$-$		\\
6371	&	\ion{Si}{ii}	&	2	&	-0.292	&	0.63	$\pm$	0.14	&		$-$		&		$-$		\\
6548	&	{\fnii}	&	1F	&	-0.318	&	29.7	$\pm$	1.6	&	23.2	$\pm$	1.1	&	27.78	$\pm$	0.89	\\
6563	&	{\hi}	&	{\ha}	&	-0.320	&	269	$\pm$	14	&	268	$\pm$	11	&	278.4	$\pm$	9.0	\\
6583	&	{\fnii}	&	1F	&	-0.323	&	89.0	$\pm$	4.8	&	73.4	$\pm$	3.2	&	85.1	$\pm$	2.8	\\
6716	&	{\fsii}	&	2F	&	-0.342	&	44.1	$\pm$	2.5	&	30.4	$\pm$	1.5	&	44.7	$\pm$	1.5	\\
6731	&	{\fsii}	&	2F	&	-0.344	&	32.9	$\pm$	1.8	&	29.6	$\pm$	1.4	&	33.6	$\pm$	1.1	\\
7318	&	{\foii}	&	2F	&	-0.418	&		$-$		&	1.57	$\pm$	0.49	&		$-$		\\
7320	&	{\foii}	&	2F	&		&				&				&				\\
7330	&	{\foii}	&	2F	&	-0.420	&		$-$		&	1.27	$\pm$	0.40	&		$-$		\\
7331	&	{\foii}	&	2F	&		&				&				&				\\
\multicolumn{4}{l}{$c$(H$\beta$)} & 1.79 $\pm$ 0.07 & 1.88 $\pm$ 0.05 & 0.74 $\pm$ 0.03 \\
\multicolumn{4}{l}{$F$(H$\beta$)$^{\rm a}$} & 0.65 $\pm$ 0.01 & 0.249 $\pm$ 0.006 & 1.97 $\pm$ 0.04 \\
     \hline                                              
    \end{tabular}
    \begin{description}
      \item[$^{\rm a}$] Flux uncorrected for reddening in units of 10$^{-14}$ erg cm$^{-2}$ s$^{-1}$.                   
    \end{description}
   \end{table*}		

%%%%%%%%%%%%%%% 

\section{The auroral {\fnii} 5755 \AA\ line in the spectra}
\label{appex:2}
In this Appendix we include sections of the spectra of all the {\hii} regions of the sample showing the quality of the measurements of the auroral {\fnii} 5755 \AA\ line, which is necessary for the determination of the electron temperature.

%%%%%%%%%%%%%%% 
\begin{figure*} 
\centering 
\includegraphics[scale=0.15]{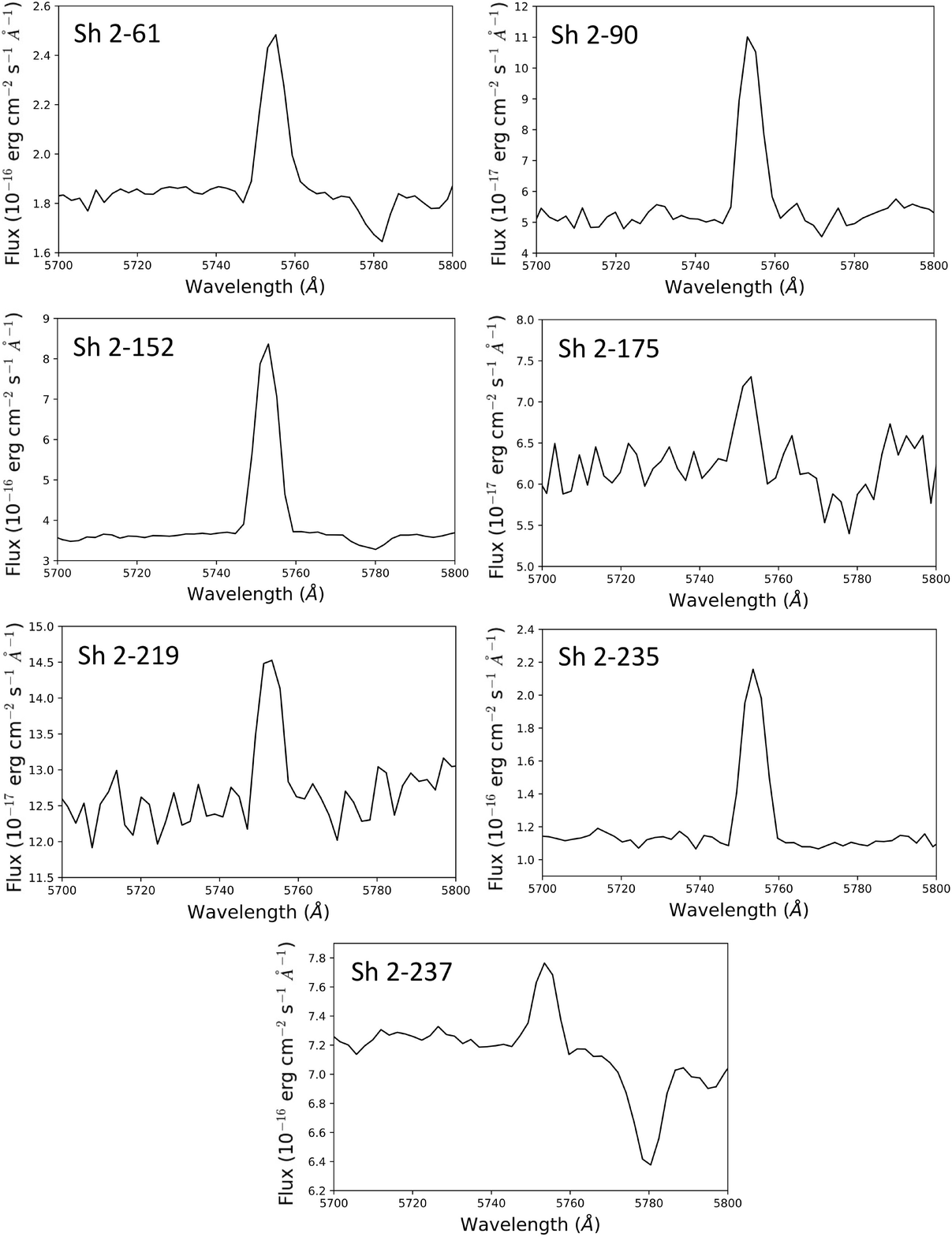} 
\caption{Section of the spectrum of Sh~2-61, Sh~2-90, Sh~2-152, Sh~2-175,  Sh~2-219, Sh~2-235 and Sh~2-237 showing the auroral {\fnii} 5755 \AA\ line. The absorption feature at 5780 \AA\ corresponds to a diffuse interstellar band.} 
\label{fig:5755_1} 
\end{figure*} 
%%%%%%%%%%%%%%%

%%%%%%%%%%%%%%% 
\begin{figure*} 
\centering 
\includegraphics[scale=0.15]{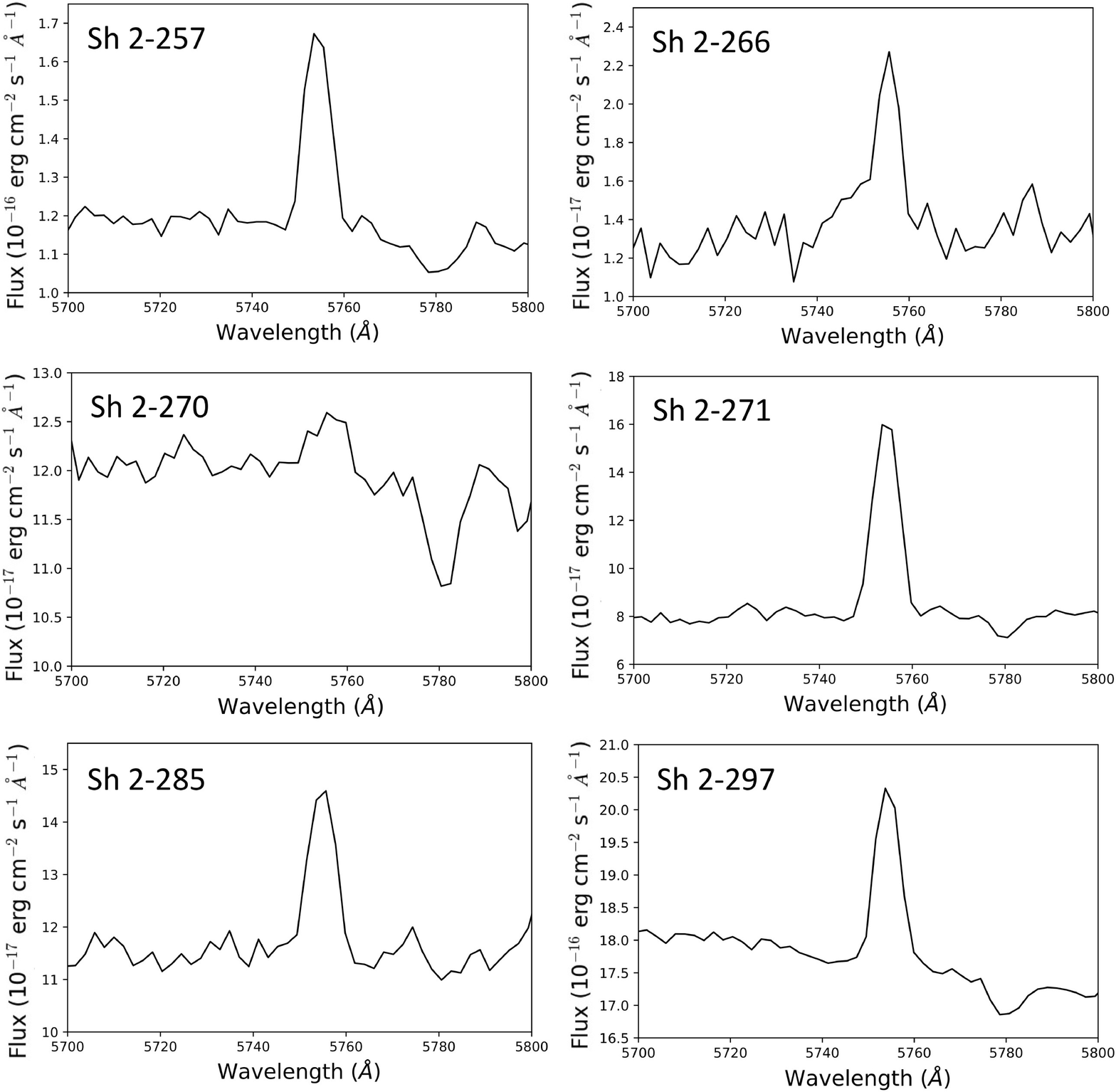} 
\caption{Section of the spectrum of Sh~2-257, Sh~2-266, Sh~2-270, Sh~2-271, Sh~2-285 and Sh~2-297 showing the auroral {\fnii} 5755 \AA\ line. The absorption feature at 5780 \AA\ corresponds to a diffuse interstellar band.} 
\label{fig:5755_2} 
\end{figure*} 
%%%%%%%%%%%%%%%

%%%%%%%%%%%%%%%%%%%%%%%%%%%%%%%%%%%%%%%%%%%%%%%%%%

% Don't change these lines 
\bsp	% typesetting comment 
\label{lastpage} 
\end{document}